\documentclass[a4paper]{article}
\usepackage{amsmath,epsfig,psfrag}

\newcommand{\nn}{\nonumber}

\newcommand{\Lam}{\Lambda}
\newcommand{\lam}{\lambda}
\newcommand{\Om}{\Omega}

\newcommand{\gam}{\gamma}

\newcommand{\alp}{\alpha}

\newcommand{\eps}{\epsilon}

\newcommand{\phbar}{{\phi}}
\newcommand{\phidot}{\dot{\phi}}
\newcommand{\Hbar}{{H}}
\newcommand{\Lbar}{{\Lambda}}
\newcommand{\diff}[2]{\frac{d #1}{d #2}}
\newcommand{\cdt}{\mathsf{t}}
\newcommand{\cdr}{\mathsf{r}}
\newcommand{\cdx}{\mathsf{x}}
\newcommand{\cdy}{\mathsf{y}}
\newcommand{\cdz}{\mathsf{z}}
\newcommand{\deriv}[2]{\frac{\partial #1}{\partial #2}}
\newcommand{\et}[1]{e^{\mbox{\small $#1$}}}
\newcommand{\clp}{{\mathcal{P}}}
\newcommand{\clr}{{\mathcal{R}}}
\newcommand{\kms}{\mathrm{\,km\,s^{-1}}}
\newcommand{\kmsmpc}{\kms{\rm\,Mpc^{-1}}}
\newcommand{\impc}{{\rm \, Mpc^{-1}}}
\newcommand{\Omm}{\Omega_{m}}
\newcommand{\del}{\delta}

\begin{document}

\noindent
{\bf\Large \textsf{Closed Universes, de Sitter Space and Inflation}}

\vspace{0.4cm}

\noindent
{\large Anthony Lasenby\footnote{e-mail: \texttt{a.n.lasenby@mrao.cam.ac.uk}}
and
Chris Doran\footnote{e-mail: \texttt{c.Doran@mrao.cam.ac.uk}}}

\vspace{0.4cm}

\noindent
Astrophysics Group, Cavendish Laboratory, Madingley Road, \\
Cambridge CB3 0HE, UK.

\begin{center}
\vspace{0.4cm}

\begin{abstract}
We present a new approach to constructing inflationary models in
closed universes.  Conformal embedding of closed-universe models in a
de Sitter background suggests a quantisation condition on the
available conformal time.  This condition implies that the universe is
closed at no greater than the 10\% level.  When a massive scalar field
is introduced to drive an inflationary phase this figure is reduced to
closure at nearer the 1\% level.  In order to enforce the constraint
on the available conformal time we need to consider conditions in the
universe before the onset of inflation.  A formal series around the
initial singularity is constructed, which rests on a pair of
dimensionless, scale-invariant parameters.  For physically-acceptable
models we find that both parameters are of order unity, so no fine
tuning is required, except in the mass of the scalar field.  For
typical values of the input parameters we predict the observed values
of the cosmological parameters, including the magnitude of the
cosmological constant.  The model produces a very good fit to the most
recent CMB (cosmic microwave background) radiation data.  The
primordial curvature spectrum provides a possible explanation for the
low-$\ell$ fall-off in the CMB power spectrum observed by WMAP
(Wilkinson anisotropy probe).  The spectrum also predicts a fall-off
in the matter spectrum at high $k$, relative to a power law. A further
prediction of our model is a large tensor mode component, with
$r\approx 0.2$.
\end{abstract}

\vspace{0.4cm}


\end{center}

\section{Introduction}

Experimental evidence for the existence of a cosmological constant has
grown dramatically over recent years.  It now seems likely that the
cosmological constant, or some form of `dark energy', is responsible
for around 70\% of the total energy density of the
universe~\cite{WMAP}.  There are two attitudes one can take towards
the dark energy.  One is that it is a relic from symmetry breaking in
a more complete theory, and as such has a field-theoretic origin.  As
is well known, this idea runs into difficulties in explaining the
current magnitude of the cosmological constant.  The second viewpoint
is the more geometric one, that the cosmological constant is related
to the large scale geometry of the universe.  This is favoured by the
fact that the energy density in the cosmological constant is currently
of the same order as that of the large scale matter density.  But this
viewpoint is silent on any mechanism that might explain the size of
$\Lam$.  Here we develop the geometric viewpoint, and suggest a novel
boundary condition that can account for the size of the cosmological
constant.

A two-dimensional space of constant negative curvature and
positive-definite signature can be conveniently visualised using the
Poincar\'{e} disk construction~\cite{brannan-cup,gap}.  The disk
represents the entire 2-dimensional space, with geodesics represented
by circle arcs that intersect the boundary at right angles.  Here we
generalise this idea to provide a Lorentzian picture of de Sitter
space.  Open, closed and flat cosmological models can be embedded into
sections of this space.  The case of closed models is of greatest
interest here, and for these we see that a natural boundary condition
presents intself.  This places the initial singularity at the midpoint
of the conformal picture of de Sitter space, which implies that the
total amount of conformal time available to the universe should equal
$\pi/2$.  For a simple dust cosmology this boundary condition picks
out a single trajectory in the $\Om_M$--$\Om_\Lam$ plane, and for the
present epoch predicts a universe that is closed at around the 10\%
level.  When coupled with knowledge of $\Om_M$, the boundary condition
therefore predicts a value of $\Om_\Lambda$ that is close to the
experimental value, in sharp contrast to predictions of $\Lam$ from
high energy physics.

To further reduce the predicted value of $\Om_\mathrm{tot}$ we
next consider an inflationary phase.  This will still be
necessary to seed the initial perturbations, and has the added
benefit of using up some conformal time before the universe
enters its matter-dominated phase. We restrict attention to the
simplest case of a single, massive scalar field $\phi$, which
ensures that the field equation for $\phi$ is linear.  This is
the most economic model that involves the least new physics.  In
order to apply our boundary condition we need to study the
evolution of the scalar field from the initial singularity,
through the inflationary region, before matching onto a standard
cosmological model.  Expanding the field equations around the
initial singularity involves a complicated iterative scheme that
can be extended to arbitrary precision.  The series is governed
by two parameters, which effectively control the degree of
inflation and the curvature.  Fixing both of these parameters to
be of order unity produces inflationary models in a closed
universe that are consistent with observation.  This runs counter
to the claim that it is difficult to obtain closed universe
inflation without excessive fine
tuning~\cite{lin03,sta96,uza03,ellis02a,ellis02}.

An attractive feature of closed-universe models is that one always has
a fixed distance scale to refer to, fixed by the size of the universe
at the time of interest.  One can use this, for example, to fix the
onset of the regime in which quantum gravity effects might be expected
to dominate.  This will be when the radius of the universe is of the
order of the Planck scale.  For the regions of parameter space of
interest here, we find that quantum gravity effects are not relevant
until well \textit{before} the onset of the inflationary region.  We
therefore have little choice but to run our equations back in time to
close to the initial singularity.  This approach has the advantage of
setting the values of the fields as the universe enters the
infaltionary regime, making the model highly predictive.

As the universe exits the inflationary region it evolves as if it had
started from an effective big-bang, with a displaced time coordinate.
By this point a sizeable fraction of the available conformal time has
been used up, which further restricts the extent to which the universe
can deviate from flatness.  The result of these effects is the
imposition of a see-saw mechanism linking the current state of the
universe and the initial conditions.  The more we increase the number
of e-folds during inflation, the smaller the value of the cosmological
constant, and vice-versa.  Furthermore, we find that $\Lam$ scales as
$\exp(-6N)$, where $N$ is the number of e-foldings during inflation,
and is in the range of 40 -- 50.  The smallness of $\Lam$ is then an
immediate consequence of this model.  With initial conditions chosen
to give the required number of e-foldings to generate the observed
perturbation spectrum, we find that the predicted universe is closed
at the level of a few percent.  Given a present value of $\Om_M$ at
around $0.3$ we single out models with $\Om_\Lam \approx 0.7$.  That
is, the conformal time constraint predicts both the degreee of
flatness of the universe and the magnitude of the cosmological
constant.

For this model to be valid it has to correctly predict the growth of
perturbations during the inflationary phase.  The primordial curvature
spectrum we predict shows a strong departure from a simple power law,
with an exponential cut-off at low wave number, and an additional
fall-off at high $k$.  The resulting CMB power spectrum is in good
agreement with the WMAP data~\cite{WMAP}, and is able to explain the
observed low-$\ell$ deficit.

We write the FRW equations as

\begin{align}
\frac{\dot{R}^2 + k}{R^2} - \frac{\Lam}{3} &= \frac{8 \pi G}{3} \rho , \\
2 \frac{\ddot{R}}{R} + \frac{\dot{R}^2 + k}{R^2} - \Lam &= - 8 \pi G P,
\end{align}
where $k=0,\pm 1$ and throughout we work in units where $c=\hbar=1$.
The scale parameter $R$ has dimensions of distance and for closed
models $R$ is the proper radius of the universe.  This is why we
prefer the symbol $R$ over the more common $a$.  Where they add
clarity, factors of $G$ are included, so that $G$ has dimensions of
$(\mbox{distance})^2$.  The ratios $\Om_M$ and $\Om_\Lam$ are defined
by
\begin{equation}
\Om_M = \frac{8 \pi G \rho}{3 H^2}, \qquad
\Om_\Lam = \frac{\Lam}{3 H^2},
\end{equation}
where $H= \dot{R}/R$.  The universe is spatially flat if
$\Om_M+\Om_\Lam=1$.

\section{Picturing de Sitter space}

\begin{figure}
\begin{center}
\includegraphics[height=6cm]{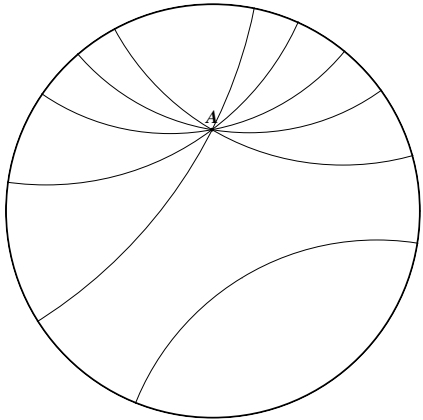}
\end{center}
\caption[dummy1]{\textit{The Poincar\'{e} disc}.
Points inside the disc represent points in a 2-dimensional
non-Euclidean (hyperbolic) space.  A set
of $d$-lines are also shown.  These are (Euclidean) circles that
intersect the unit circle at right angles.  Given a $d$-line and point
$A$ not on the line, one can find an infinite number of lines through
$A$ that do not intersect the line.}
\label{fig2}
\end{figure}

\begin{figure}
\begin{center}
\includegraphics[height=5cm,angle=-90]{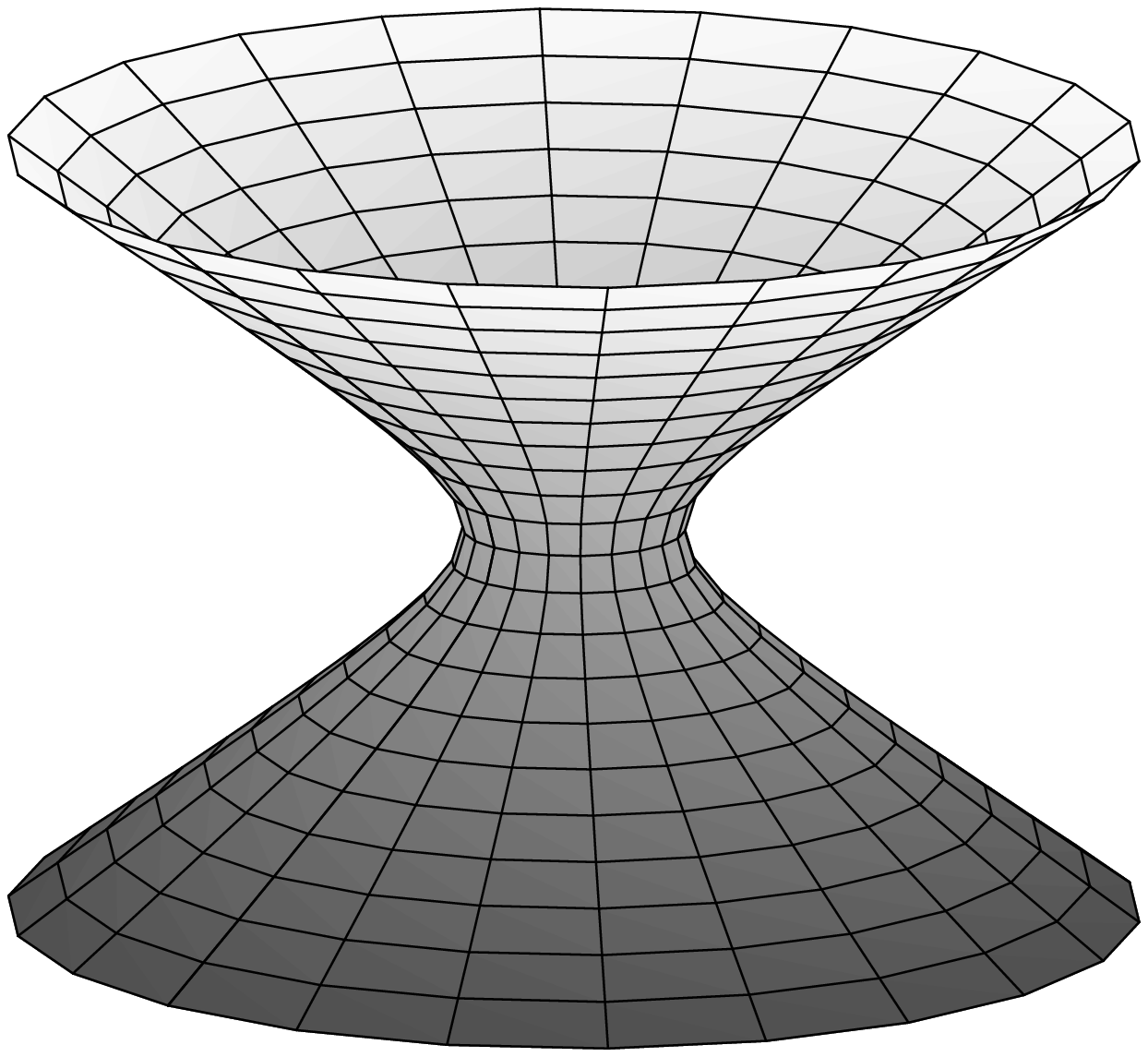}
\includegraphics[height=6cm,angle=-90]{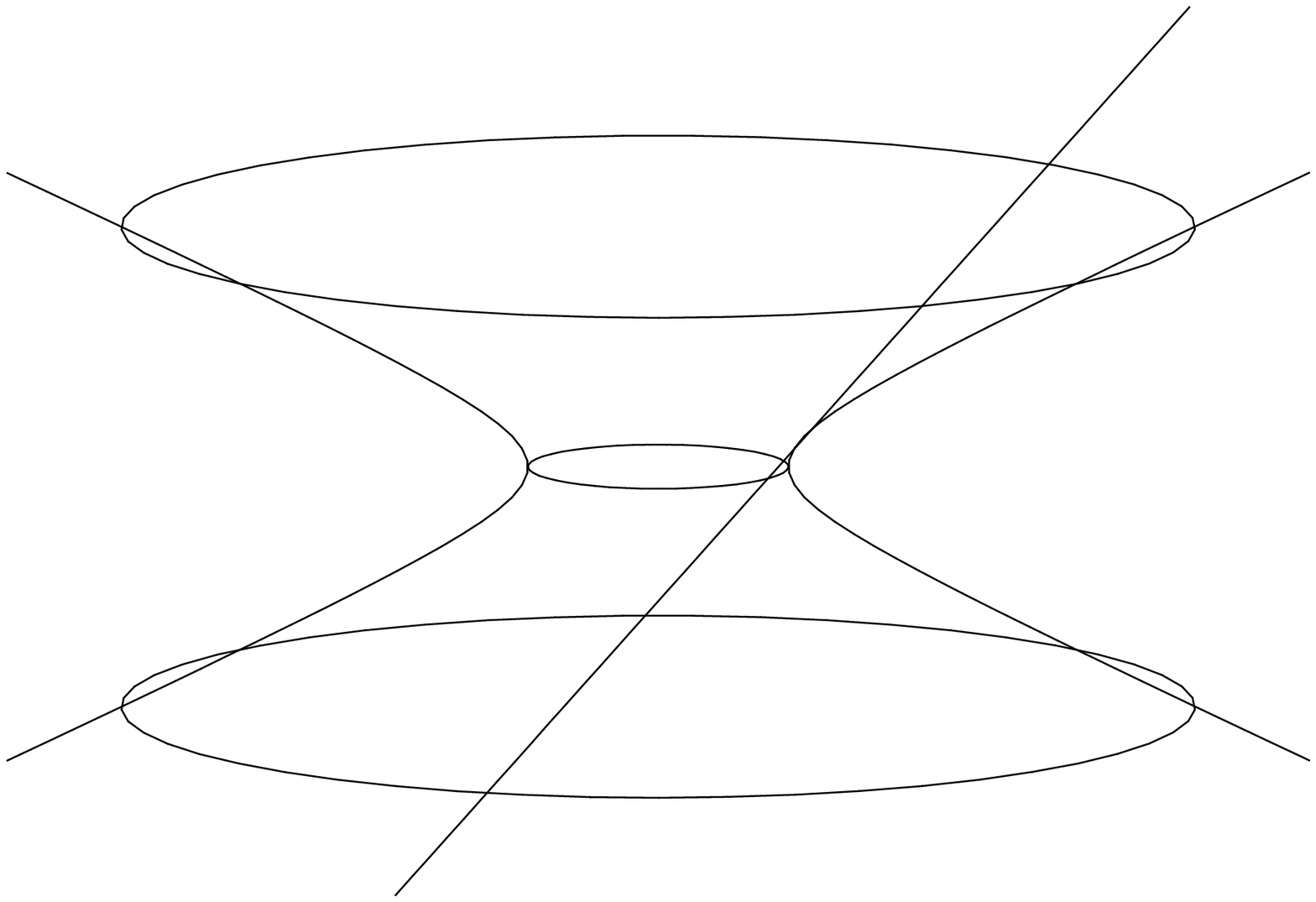}
\end{center}
\caption[dummy1]{\textit{Two-dimensional de Sitter Space}.  The
  timelike direction is vertical, and spatial sections are closed.
  The right-hand diagram shows a null geodesic, which is a straight
  line in the embedding space.}
\label{fig4}
\end{figure}

A de Sitter space is a space of constant negative curvature.  As such,
it forms the Lorentzian analogue of the non-Euclidean geometry
discovered by Lobachevskii and Bolyai~\cite{brannan-cup,gap}.
Two-dimensional non-Euclidean geometry has an elegant construction in
terms of the Poincar\'{e} disc, in which geodesics are represented as
`$d$-lines' --- circles that intersect the disc boundary at
right-angles (see figure~\ref{fig2}).  Here we provide a similar
picture for 2-dimensional de Sitter space, which is sufficient to
capture all of the key features of the geometry~\cite{phisky}.  We
start with an embedding picture, representing de Sitter space as the
2-surface defined by
\begin{equation}
T^2 - X^2 - Y^2 = -a^2,
\end{equation}
where $(T,X,Y)$ denote coordinates in a space of signature $(1,2)$ and
$a$ is a constant.  The resulting surface is illustrated in
figure~\ref{fig4}.  The figure illustrates some key features of the
global de Sitter geometry.  In particular, spatial sections are
closed, whereas the timelike direction is open.  So de Sitter space
describes a closed universe that lasts for infinite time.  One can
also set up local coordinate patches for which spatial sections are
flat or open, but these coordinates are not global and only cover part
of the full manifold.  Null geodesics are straight lines formed by the
intersection of the surface and a vertical plane a distance $a$ from
the timelike axis.  Despite the fact that the space is spatially
closed, the furthest a photon can travel is half of the way round the
universe.

\begin{figure}
\begin{center}
\includegraphics[height=6cm]{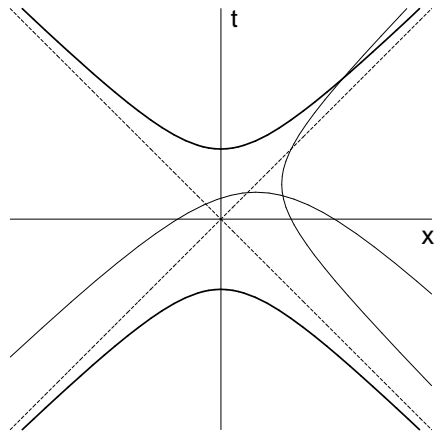}
\end{center}
\caption[dummy1]{\textit{The Lorentzian view of two-dimensional de
Sitter space}.  The boundary is defined by two hyperbolae (shown with
thick lines).  All geodesics through the origin are straight lines,
and null geodesics are always straight lines at $45^\circ$.  Two
further geodesics, one spacelike and one timelike, are also shown.
These are hyperbolae which do not pass through the origin.  The
timelike geodesic intersects the boundary in such a way that the two
tangent vectors have vanishing Lorentzian inner product.}
\label{fig6}
\end{figure}

To establish an analogous picture to the Poincar\'{e} disc for de
Sitter space, we start by considering the spatial section at $T=0$.
This section is a ring of radius $a$, which is mapped onto a straight
line in a $(1,1)$ Lorentzian space via a stereographic projection.
Null geodesics from this section are now represented as $45^\circ$
straight lines in Lorentzian space.  Since geodesics from opposite
points on the ring meet at infinity, we arrive at a boundary in the
timelike direction defined by a hyperbola.  This construction provides
us with a Lorentzian view of de Sitter geometry.  Timelike geodesics
in de Sitter space are represented by hyperbolae that intersect the
boundary at a right-angle (see figure~\ref{fig6}).  Here
`right-angle' is defined in its Lorentzian sense (the tangent
vectors have vanishing Lorentzian inner product).  This view has the
convenient feature that geodesics are lines of constant Lorentzian
distance from some fixed point.  The metric associated with this view
of de Sitter geometry has the conformal structure
\begin{equation}
ds^2 = \frac{a^4}{(a^2+\cdx^2-\cdt^2)^2}(d\cdt^2 - d\cdx^2),
\end{equation}
which makes it clear that null geodesics must remain as straight lines
in the $\cdx$--$\cdt$ plane.

There are many fascinating geometric structures associated with de
Sitter geometry, many of which mirror those of non-Euclidean geometry.
For example, one can always find a reflection that takes any point to
the origin~\cite{anl-confcos}.  One can then prove a number of results
at the origin, where the geodesics are all straight lines, and the
results are guaranteed to hold at all points.  A similar approach can
be applied to anti-de Sitter space, except now the diagram is rotated
through $90^\circ$, as it is the timelike direction that is formed
from a stereographic unwrapping of a circle.

\section{Cosmological models and the de Sitter end state}

\begin{figure}
\begin{center}
\includegraphics[width=5cm]{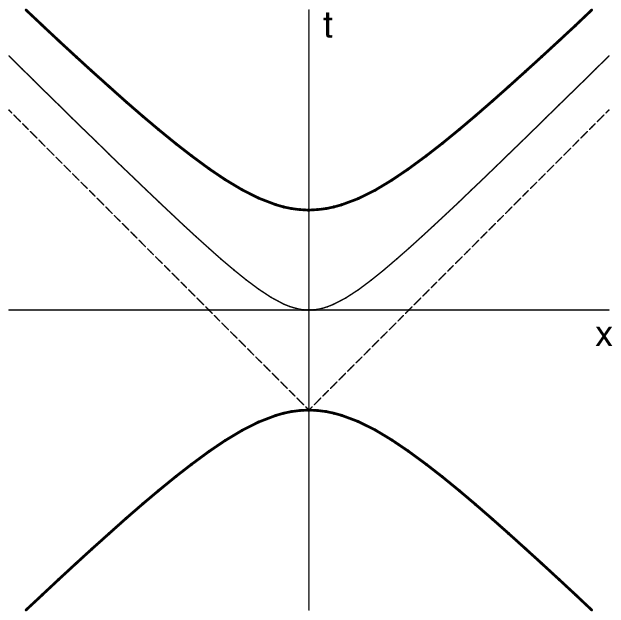}
\includegraphics[width=5cm]{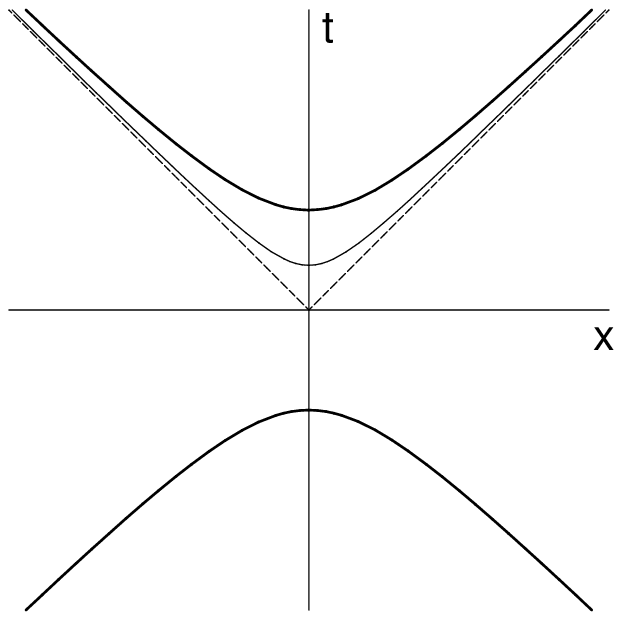}
\end{center}
\caption[dummy1]{\textit{Flat and open cosmological models}.  de
Sitter space contains sections representing both flat and open
spacetimes.  A flat spacetime (left) consists of the space contained
inside a light-cone located on past infinity.  An open spacetime fits
inside the lightcone from the origin.  Both pictures can be used to
illustrate cosmological models, with the initial singularity
represented by a hyperbolic spacelike surface.  An example of these
surfaces is shown on each diagram.}
\label{fig7}
\end{figure}

All cosmological models are conformally flat, and can all be
interchanged via conformal transformations.  Furthermore, all models
containing a cosmological constant, and which do not recollapse, will
tend towards a de Sitter end.  Such models should fit neatly within
the conformal diagram of figure~\ref{fig6} with future infinity
represented by the upper hyperbolic boundary.  There are three cases
to consider, for open, closed and flat cosmologies.  In this paper we
are interested in closed cosmologies, but before looking at these it
is informative to consider the other cases.  A flat section of de
Sitter space corresponds to a region contained within a light cone
from a point located on the boundary at past infinity (see
figure~\ref{fig7}).  Surfaces of constant cosmic time are represented
as hyperbolae, any one of which can then be chosen to represent the
initial singularity.  Similarly, an open section of de Sitter space is
represented by the area inside a light-cone from a point in the middle
of the de Sitter picture.  A more general open $\Lam$-cosmology will
have the initial singularity located on a spacelike hyperbola.  (See
Ratra and Peebles~\cite{rat95} for more details of inflation in open
cosmologies.)

The diagrams for flat and open cosmologies make it clear that one is
not employing the full de Sitter geometry in a symmetric manner.
Closed models, on the other hand, have a more natural embedding, with
the initial singularity represented by a spacelike surface in the
conformal diagram of figure~\ref{fig6}.  Here we begin to see a
natural boundary condition emerging.  Suppose that we insist that
$\cdt=0$ represents the initial singularity, so that the entire
history of the universe can be conformally mapped into the top half of
the de Sitter diagram.  (This does not imply that the actual universe
is static, it is merely a statement about its conformal
representation.)  This is certainly the most natural, symmetric place
to locate the singularity.  An immediate consequence is that any
photon emitted from the initial singularity may travel a maximum of
precisely one-quarter of the way round the universe in the entire
future evolution of the universe.  An alternative way of saying this
is that the past horizon projected back to $\cdt=0$ should cover half
of the de Sitter geometry.

Given an equatorial angle $\phi$ on a 3-sphere, a photon travelling
round the equator will satisfy
\begin{equation}
\frac{d\phi}{dt} = \frac{1}{R},
\end{equation}
where $t$ is cosmic time.  Traversing the entire universe corresponds
to running from $0\leq \phi \leq 2\pi$.  If a photon is only to travel
one-quarter of the way round we therefore require that
\begin{equation}
\int_0^\infty \frac{dt}{R} = \int_0^\infty \frac{dR}{R(-1 + \Lam R^2/3 +
  8\pi \rho R^2/3)^{1/2}} = \pi/2.
\label{etacond}
\end{equation}
At this point is is useful to define the conformal time $\eta$ by
\begin{equation}
\eta = \int_0^t \frac{dt'}{R(t')}.
\label{defeta}
\end{equation}
Note that for closed models the conformal time $\eta$ is not the same
as the time-like conformal coordinate $\cdt$ (see
appendix~\ref{appa}).  Our boundary condition now states that the end
of the universe corresponds to a total passing of $\pi/2$ units of
conformal time.

The constraint $\eta(\infty)=\pi/2$ can be arrived at through an
alternative route that works entirely within the conformal
representation of cosmological models, with the line element
written in the form
\begin{equation}
ds^2 = \frac{1}{f^2} \bigl( d\cdt^2 - d\cdx^2 - d\cdy^2 - d \cdz^2
\bigr).
\end{equation}
The precise form of $f$ depends on the spatial topology. In
appendix~\ref{appa} we show that the three cases reduce to
\begin{align}
f &= \frac{\cdt}{R \sin(\eta)} =
g \left( \frac{2 \lam \cdt}{\lam^2 + \cdr^2 - \cdt^2} \right) \,
\frac{\cdt}{\lam} &
\mbox{closed} \nn \\
f &= \frac{2 \lam}{R} & \mbox{flat}  \\
f &=\frac{\cdt}{R \sinh(\eta)} = \bar{g} \left(\frac{2 \lam
  \cdt}{\lam^2 + \cdt^2 - \cdr^2} \right)  \,
\frac{\cdt}{\lam} & \mbox{open} \nn
\end{align}
where $\cdr^2 = \cdx^2 + \cdy^2 + \cdz^2$, $\lambda$ is a constant
with dimensions of distance, and the forms of $g$ and $\bar{g}$
depends on the model for the matter content.  The coordinate ranges
are as illustrated in figures~\ref{fig6} and~\ref{fig7}. A de Sitter
space centred on $t=0=\cdt$ has the conformal line element
\begin{equation}
ds^2 = \frac{12 \lam^2}{\Lam(\lam^2 + \cdr^2 - \cdt^2)}  \bigl(
d\cdt^2 - d\cdx^2 - d\cdy^2 - d \cdz^2 \bigr).
\end{equation}
Clearly the only solutions that have any chance of matching onto
this final state are those for a closed universe.  Furthermore, the
function $g$ must satisfy
\begin{equation}
\lim_{\chi \mapsto \infty} g(\chi) = \left( \frac{\lam^2 \Lam
}{3}\right)^{1/2} \frac{1}{\chi}
\end{equation}
where
\begin{equation}
\chi = \frac{2 \lam \cdt}{\lam^2 + \cdr^2 - \cdt^2} = \tan(\eta).
\end{equation}
But since
\begin{equation}
g = \frac{\lam}{R \sin(\eta)}
\end{equation}
we must then have $R \cos(\eta)$ tending to a constant at large
times. This is only possible if $\eta$ tends to $\pi/2$, recovering
our earlier boundary condition.  This derivation is instructive in
that it reveals how the constraint can be imposed as a straightforward
boundary condition on a differential equation.  In this case, the
equation for $g(\chi)$ is
\begin{equation}
\chi^2(1+\chi^2) \left( \frac{dg}{d\chi}\right)^2 + \frac{d}{d\chi}
(g^2 \chi) = \frac{8 \pi G \lam^2 \rho}{3} + \frac{\lam^2 \Lam}{3}.
\end{equation}
The task then is to solve this equation subject to the boundary
condition that $g$ falls off as $1/\chi$ for large $\chi$.  Viewed
this way the constraint can be thought of as a `quantisation
condition' applied as the universe is formed, which one might hope
would emerge from a quantum theory of gravity.

\begin{figure}
\begin{center}
\psfrag{Omm}{\small $\Omega_M$}
\psfrag{Oml}{\small $\Omega_\Lambda$}
\includegraphics[height=4.5cm]{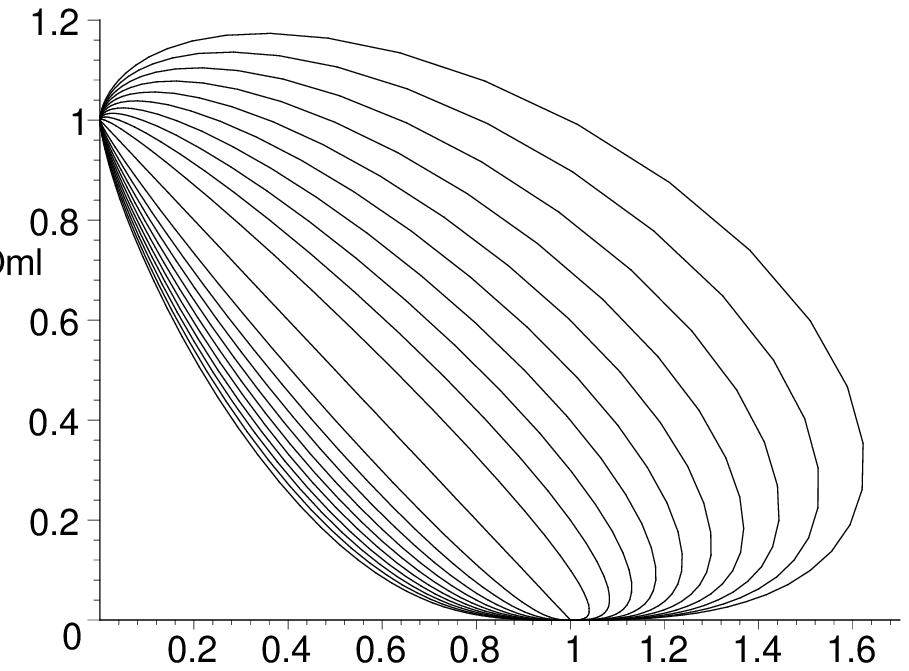}
\includegraphics[height=4.5cm]{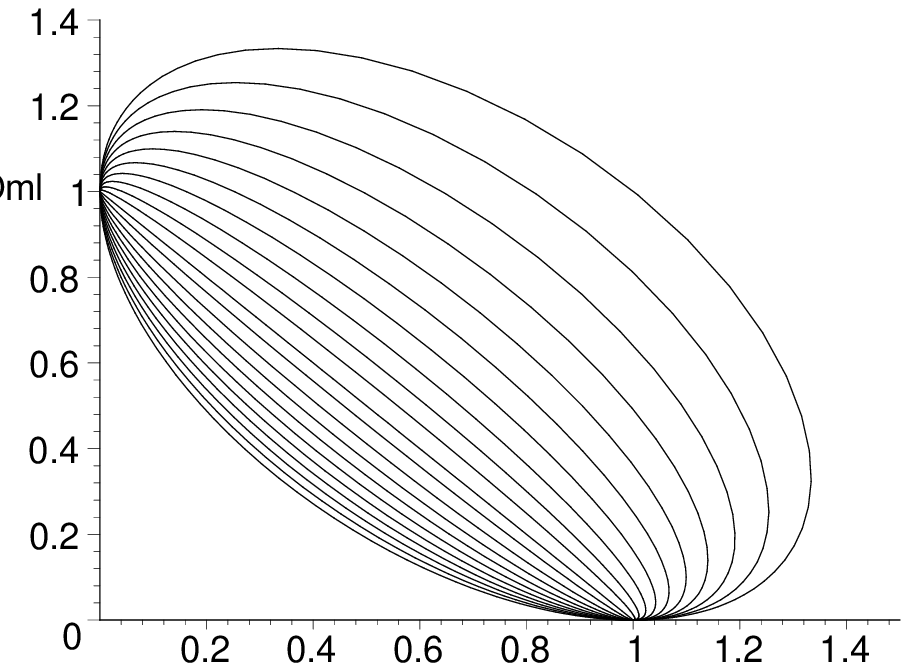}
\end{center}
\caption[dummy1]{Evolution curves in the $\Om_M$--$\Om_\Lam$
plane.  The left-hand plot is for dust, and the right-hand plot is for
radiation.  In both cases the curves converge to $\Om_\Lam=1$,
representing a late-time de Sitter phase.}
\label{fig1}
\end{figure}

In order to understand the implications of our boundary condition, we
first write the FRW equations in the form
\begin{equation}
\frac{d \Om_\Lam}{d \Om_M} = \frac{\Om_\Lam \bigl( (1+3 \gam) \Om_M -2
(\Om_\Lam -1) \bigr)}{\Om_M \bigl( (1+3 \gam)(\Om_M-1) - 2 \Om_\Lam
\bigr)},
\label{ommoml}
\end{equation}
where $\gam$ encodes the equation of state in the form
\begin{equation}
P = \gam \rho.
\end{equation}
Equation~\eqref{ommoml} defines a series of flow lines in the
$\Om_M$--$\Om_\Lam$ plane, and also highlights the significance of the
points $(1,0)$ and $(0,1)$, which are attractors of the flow lines.
This behaviour is illustrated in figure~\ref{fig1}, which shows a
series of flow lines starting from $\Om_M=1$, $\Om_\Lam=0$.

An alternative version of the equations, more useful in describing the
universe since recombination, is to assume that the matter density is
made up of decoupled dust and radiation.  Writing the densities of
these as $\rho_m$ and $\rho_r$ respectively, we introduce the
quantities
\begin{equation}
\Om_m = \frac{8 \pi G \rho_m}{3 H^2},
\quad \mbox{and} \quad
\Om_r = \frac{8 \pi G \rho_r}{3 H^2} .
\end{equation}
Since the matter and radiation are decoupled, both stress-energy
tensors satisfy separate conservation laws, which reduce to
\begin{equation}
\rho_m R^3 = \mbox{constant},
\quad \mbox{and} \quad
\rho_r R^4 = \mbox{constant}.
\end{equation}
For this case we find that the equations governing $\Om_\Lam$, $\Om_m$
and $\Om_r$ can be solved exactly, with the solution governed by two
arbitrary constants.  We label these $\alp$ and $\beta$, where
\begin{equation}
\alpha = \frac{\Om_m^2 \Om_\Lam}{(\Om_m + \Om_r + \Om_\Lam -1)^3}  =
\frac{(8 \pi G \rho_m R^3)^2 \Lam}{27},
\label{defalp}
\end{equation}
and
\begin{equation}
\alpha \beta =  \frac{\Om_r \Om_\Lam}{(\Om_m + \Om_r + \Om_\Lam -1)^2} =
\frac{8 \pi G \rho_r R^4 \Lam}{9}.
\label{defgam}
\end{equation}
Present knowledge of $\Om_m$, $\Om_r$ and $\Om_\Lam$ fixes $\alpha$
and $\beta$, and so picks out a unique curve.  The curve will be
valid back to last scattering, beyond which the equation of state
becomes more complicated.

The diagrams in figure~\ref{fig1} highlight the fact that as the
universe evolves the flow lines refocus around the spatially flat
case, $\Om_M+\Om_\Lam=1$.  One can show that, for a large range of
initial conditions, by the time we reach the current value of $\Om_M
\approx 0.3$ most models are not far off spatial flatness.  This
prediction contrasts with models without a cosmological constant,
where any slight deviation from the critical density in the early
universe is scaled enormously by the time we reach the present epoch,
implying that the parameters in the early universe are highly
fine-tuned.  One can see the problem straightforwardly by writing
\begin{equation}
\kappa = \Om_m + \Om_r + \Om_\Lam - 1.
\end{equation}
The parameter $\kappa$ measures the deviation from spatial flatness,
and satisfies
\begin{equation}
\frac{\dot{\kappa}}{\kappa} = H(\Om_m + 2 \Om_r - 2\Om_\Lam).
\end{equation}
If $\Om_\Lam=0$ the right-hand side is always positive, and nothing
prevents the continued growth of $\kappa$.  The presence of a $\Lam$
term changes this completely.  At some finite time $2\Om_\Lam$ will
overtake the matter terms, and $\kappa$ is refocused back to
$\kappa=0$. The presence of a cosmological constant therefore goes
someway to solving the flatness problem on its own, without having to
invoke inflation.

\begin{figure}
\begin{center}
\psfrag{Omm}{$\Omega_M$}
\psfrag{Oml}{$\Omega_\Lambda$}
\includegraphics[height=6cm]{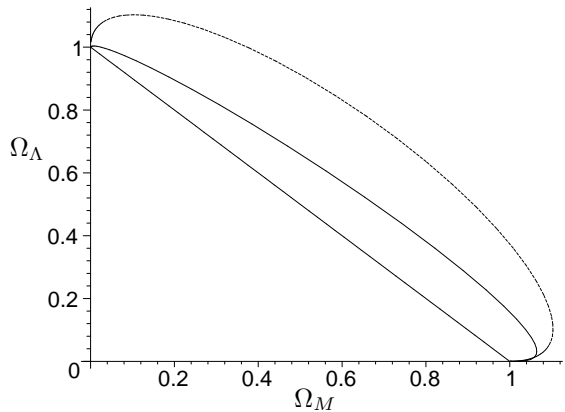}
\end{center}
\caption[dummy1]{\textit{Critical paths as predicted by the de Sitter
embedding}.  The solid line represents a matter-dominated universe,
and the broken line shows radiation, for comparison.  The straight
line is the critical case of spatial flatness.  By the time we reach
$\Om_M \approx 0.3$ the universe has been driven close to critical
density.}
\label{fig8}
\end{figure}

For the case of a universe filled with non-interacting dust and
radiation we can write the boundary condition of
equation~\eqref{etacond} as
\begin{equation}
\int_0^\infty \frac{dx}{( \beta x^4 + x^3 - x^2 + \alpha)^{1/2}} = \pi
/2.
\label{etacond2}
\end{equation}
This equation is therefore an eigenvalue problem for the dimensionless
ratios $\alpha$ and $\beta$.  For the case of dust ($\beta=0$) the
solution turns out to be $\alpha \approx 40.468$.  Pinning down a
value of $\alp$ and $\beta$ picks out a unique trajectory in the
$\Om_M$--$\Om_\Lam$ plane.  For the straightforward cases of dust
($\beta=0$) and radiation ($\alp=0$) we arrive at the two curves shown
in figure~\ref{fig8}.  As the universe is expected to be matter
dominated for most of its history, the solid line in figure~\ref{fig8}
is the more physically relevant one.  Taking the present-day energy
density to be around $\Om_M = 0.3$, we see that $\Om_\Lam=$ is
predicted to be around $\Om_\Lam \approx 0.83$.  That is, a universe
that is closed at around the 10\% ratio.  Such a prediction is
reasonably close to the observed value, though it is ruled out by the
most recent experiments~\cite{WMAP}.  In order to improve the
prediction, we need to use up a greater fraction of the conformal time
before we enter the matter-dominated phase.  Such a process is also
necessary to solve the horizon problem, and the simplest means of
achieving this is via an inflationary phase.

\section{Scalar fields and inflation}
\label{Sinflat}

Much effort has gone into constructing quantum field theories in de
Sitter space~\cite{che68,bir-quant,all85,dan02}.  There are two
reasons why this is more complicated that in Minkowski space: the
closed spatial sections give rise to a quantization condition; and the
presence of event horizons give rise to ambiguities in the definition
of the vacuum.  Similar issues can arise when considering an evolving
closed universe and will be relevant for a complete treatment of the
growth of perturbations seeded by a scalar field.  In this section we
will ignore such issues, delaying a discussion of the quantization
procedure until section~\ref{S-curv}.  Our model for the matter
therefore consists simply of a real, time-dependent, homogeneous
massive scalar field.  For a range of initial conditions this system
shows the expected inflationary behaviour.  The equations for this
model are
\begin{equation}
\dot{H} + H^2 - \frac{\Lambda}{3} + \frac{4 \pi G}{3} ( 2
\dot{\phi}^2 - m^2 \phi^2) = 0
\label{sceq1}
\end{equation}
and
\begin{equation}
\ddot{\phi} + 3 H \dot{\phi} + m^2 \phi = 0.
\label{sceq2}
\end{equation}
Here $\phi$ has dimensions of $(\mbox{length})^{-1}$, which is the
convention widely adopted in the cosmology literature (though this
does differ from the relativistic quantum mechanics
literature~\cite{gre-rqm}).  As stated in the introduction, it is
convenient to keep in explicit factors of $G$, to keep the units
clear. For closed models the scale factor is given explicitly by
\begin{equation}
\frac{1}{R^2} = \frac{4 \pi G}{3} (\dot{\phi}^2 + m^2  \phi^2) - H^2 +
\frac{\Lambda}{3}.
\label{curv}
\end{equation}

Given a solution to the pair of equations~\eqref{sceq1}
and~\eqref{sceq2}, a new solution set is generated by scaling with a
constant $\sigma$ and defining
\begin{equation}
H'(t) = \sigma \Hbar(\sigma t), \quad \phbar' (t) =
 \phbar(\sigma t), \quad m' =  \sigma m, \quad \Lbar' =
\sigma^2 \Lbar.
\label{scltrf}
\end{equation}
This scaling property is valuable for numerical work, as a range of
situations can be covered by a single numerical
integration.  Furthermore, many physically interesting quantities turn
out to be invariant under changes in scale.  These include the
conformal time $\eta$, which is easily confirmed to be scale invariant
from equation~\eqref{defeta}.  Invariant quantities such as $\eta$
turn out to be extremely helpful when considering the initial
conditions as the universe emerges from the big bang.  The scaling
property does not survive quantisation, however, so has to be employed
carefully when considering vacuum fluctuations.

The initial conditions for equations~\eqref{sceq1} and~\eqref{sceq2}
are usually set at the start of the inflationary period, where they
are viewed as arising from some form of quantum gravity interaction.
But in order to apply our boundary condition we need to track the
equations right back to the initial singularity, as this is the only
way that we can keep track of the total conformal time that has
elapsed.  As we argue below, evolving the inflationary equation
backwards in time is entirely justified, as we do not expect quantum
gravity to play a role until much earlier in the history of the
universe.  Inflation on its own does not eliminate singularities from
cosmology~\cite{brand01}.  Our approach has the added benefit that the
conditions at the start of inflation are fully determined by a pair of
parameters, making the model highly predictive.

As the universe emerges from the big bang the dominant behaviour of
$H$ is to go as $1/(3t)$.  Equation~\eqref{sceq1} then implies that
$\phi$ must contain a term going as $\ln(t)$.  But this in turn
implies that $H$ must also contain a term in $t\ln(t)$, in order to
satisfy equation~\eqref{sceq2}.  Working in this manner we conclude that
a series expansion in terms of $\ln(t)$ is required to describe the
behaviour around the singularity.  At this point it is convenient to
define the dimensionless variables
\begin{equation}
u = \frac{t}{t_p}, \qquad \mu = \frac{m}{m_p}
\end{equation}
with $t_p$ and $m_p$ the Planck time and mass respectively.  The
series expansion about the singularity at $t=0$ can now be written
\begin{align}
\Hbar(u) &= \frac{1}{t_p}\sum_{i=0}^\infty H_i(u) \ln^i(u) \\
\phbar(u) &= \frac{1}{l_p} \sum_{i=0}^\infty \phi_i(u) \ln^i(u) ,
\end{align}
which ensures that the expansion coefficients are all dimensionless.
Substituting these into the two evolution equations, and setting each
coefficient of $\ln(u)$ to zero, we establish that
\begin{align}
H_1 &= -u \diff{H_0}{u} - u H_0^2 + \frac{u \Lbar}{3} - \frac{8
\pi u }{3 } \left(\diff{\phi_0}{u} \right)^2 - \frac{16 \pi \phi_1}{3 }
\diff{\phi_0}{u} \nn \\
& \quad - \frac{8 \pi \phi_1^2}{3 u} + \frac{4 \pi \mu^2 u
\phi_0^2}{3} ,
\end{align}
with further algebraic equations holding for $H_2, \phi_2$, $H_3,
\phi_3$, and so on.  So, by specifying $H_0$, $\phi_0$ and $\phi_1$,
all the remaining terms in the solution are fixed.  The aim then is to
choose the three input functions to ensure that successive terms in the
series get progressively smaller.  This turns out to provide just the
right number of equations to specify all coefficients, save for two
arbitrary coefficients in $\phi_0$.  The end result is a series
controlled entirely by a pair of arbitrary constants: the expected
number of degrees of freedom once we have fixed the singularity to
$t=0$.  In order to generate curvature it turns out that the input
functions need to be power series in $u^{1/3}$, which ensures that the
scale factor goes as $u^{1/3}$ at small times.  The series solution is
only required to find suitable initial conditions for numerical
evolution, so only the first few terms are required.  Expanding up to
order $u^{5/3}$ we find that
\begin{align}
H_0 &= \frac{1}{3u} + \frac {32\sqrt {3\pi }}{27}  b_4 u^{1/3}
\nn \\
& \quad + \left( \frac {2 \mu^2 }{81} + \frac{\Lbar}{3} +
\frac{4 \pi}{3} \mu^2 b_0^2+  \frac {4 \sqrt{3 \pi}}{27} {\mu}^{2}
b_0 \right) u - \frac{6656 \pi b_4^2 }{891} u^{5/3}, \\
\phi_0 &= b_0  + b_4 u^{4/3} - \frac {118 \sqrt {3 \pi} b_4^2}{99}
u^{8/3} \nn \\
& \quad
- \frac {1}{1296 \pi} \left( 11 \sqrt {3 \pi} \mu^{2}
- 54 \sqrt {3 \pi} \Lbar - 216 \sqrt {3}\, \pi^{3/2} \mu^2 b_0^2 + 36
\pi \mu^2 b_0 \right) u^2
\end{align}
and
\begin{equation}
\phi_1 = - \sqrt{\frac{1}{12 \pi}} -{\frac {\mu^2}{216 \pi}} \left
(-\sqrt {3 \pi } + 36  \pi b_0\right ) u^2 .
\end{equation}
Under scaling the three key parameters in the model transform as
\begin{equation}
\mu \mapsto \sigma \mu, \qquad b_0 \mapsto b_0, \qquad b_4
\mapsto \sigma^{4/3} b_4.
\label{scl1}
\end{equation}
The scaling transformation for $b_4$ is entirely as expected, given
that it is the coefficient of $u^{4/3}$ in the series for $\phi_0$.
It follows that the quantity $b_4/\mu^{4/3}$ is scale invariant.

\begin{figure}
\begin{center}
\includegraphics[width=6cm,angle=-90]{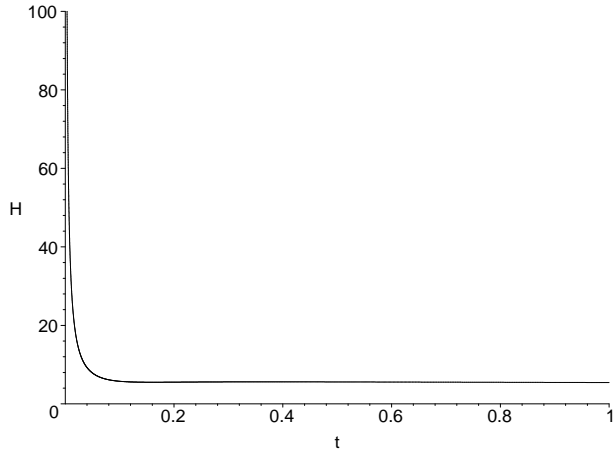}
\end{center}
\caption[dummy1]{\textit{Hubble function entry into the inflationary
regime.}  The Hubble function emerges from the big bang going as
$1/3t$.  As $H$ falls it quickly enters the inflationary region,
during which it falls linearly.  Time $t$ is measured in units of
$t_p$, and $H$ in units of $t_p^{-1}$.  The input parameters were set
to $b_0=2.48$, $b_4=-0.51$ and $m=m_p$ ($\mu=1$).  These values of
$b_0$ and $b_4/\mu^{4/3}$ are illustrative of physically realistic
models.  The cosmological constant term has no effect in this region
and can be set to zero.}
\label{fHfrom0}
\end{figure}

The plot in figure~\ref{fHfrom0} illustrates the general behaviour
around the initial singularity.  As the universe emerges from the big
bang the energy density in the scalar field is dominated by the
$\dot{\phi}$ term, and the field behaves as if it is massless.  It
follows that $H$ initially falls as $1/(3t)$.  But once $H$ has fallen
sufficiently far we enter a region in which $m^2 \phi^2$ starts to
dominate over $\dot{\phi}^2$.  These are suitable initial conditions
for the universe to enter an inflationary phase.  By varying $b_0$,
$b_4$ and $\mu$ we control both the values of the fields as we enter
the inflationary period, and how long the inflationary period lasts.
The cosmological constant plays no significant role in this part of
the evolution. The dynamics displayed in this figure is quite robust
over a range of input parameters.

The fact that $b_4$ controls the curvature can be seen from
equation~\eqref{curv} which, to leading order, yields
\begin{equation}
\frac{R}{l_p} = \frac{1}{\mu }\left( \frac{2187}{12544 \pi}
\right)^{1/4} \left( - \frac{\mu^{4/3}}{b_4} \right)^{1/2} (\mu
u)^{1/3} + \cdots.
\label{Rnear0}
\end{equation}
Clearly, the arbitrary constant $b_4$ must be negative for a closed
universe.  The terms on the right-hand side of equation~\eqref{Rnear0}
are all scale invariant, apart from the first factor of $\mu^{-1}$.
From figure~\ref{fHfrom0} we can see that, for the typical values of
$b_0$ and $b_4/\mu^{4/3}$ of interest, the onset of inlation
corresponds to $\mu u \approx 0.1$.  As we discuss below, in order to
generate perturbations consistent with observation the scalar field
must have a low mass, with $\mu$ of the order of $10^{-6}m_p$.  It
follows that the onset of inflation occurs at a time of around $10^5$
Planck times.  Furthermore, the radius of the universe at the onset of
inflation is given approximately by
\begin{equation}
R \approx \frac{0.2}{\mu} l_p,
\end{equation}
and is of order $10^5$ Planck lengths.  Inflation therefore starts at
an epoch well into the classical regime.  Quantum gravity effects
would be expected to be relevant when the radius of the universe is of
the order of the Planck scale, which occurs when $u \approx \mu^2$ and
is well before any inflationary period (for physical values of $\mu$).
We are therefore quite justified in running the evolution equations
back past the inflationary regime, and right up to near the initial
singularity.  It is only when $R=l_p$ that the equations will break
down, and we would look to quantum gravity to explain the formation of
the initial, Planck-scale sized universe.  We might hope here that
quantum gravity will provide a probability distribution for the two
classical parameters controlling the evolution of the scalar field.
Furthermore, we can argue more strongly that we \textit{have} to run
the equations backwards in time to well before the start of inflation
before reaching an epoch where new physics could be expected to enter
the problem.

As the universe is described by a 3-sphere of radius $R$,
the total volume of the universe is given to leading order by
\begin{equation}
V = 2 \pi^2 \left( \frac{2187}{12544 \pi} \right)^{3/4}
\frac{l_p^3}{(-b_4)^{3/2}} u + \cdots.
\end{equation}
Following from this, an interesting calculation we can perform in a
closed universe is to find the total energy contained within it in the
scalar field.  By integrating the energy density we find that
\begin{equation}
E_{\mbox{tot}} = \frac{\pi}{12} \left(\frac{2187}{12544 \pi}
\right)^{3/4} \frac{1}{(-b_4)^{3/2}} \frac{\hbar}{t} + \cdots \approx 0.03
\left( \frac{-\mu^{4/3}}{b_4} \right)^{3/2}  \frac{\hbar}{\mu^2 t}.
\label{Etot0}
\end{equation}
We return to a discussion of this equation in section~\ref{Sdisc}.

\begin{figure}
\begin{center}
\includegraphics[height=9cm,angle=-90]{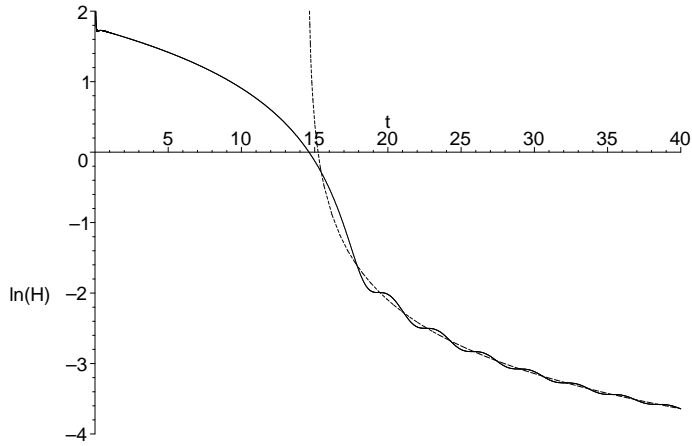}
\end{center}
\caption[dummy1]{\textit{Hubble function exit from the inflationary
    regime.}  As the inflationary era ends the density and pressure
    start oscillating around values for a matter-dominated cosmology.
    The effective singularity for this dust cosmology is displaced
    from $t=0$.  The broken line plots the natural log of
    $2/3(t-14.56)$.  The parameters in this model are $b_0=2.48$,
    $b_4=-0.51$ and $m=m_p$. }
\label{fHout}
\end{figure}

The typical behaviour as we exit the inflationary region is shown in
figure~\ref{fHout}.  The end of inflation is characterised by $\Hbar
\approx \mu$.  Beyond this point the scalar field enters an
oscillatory phase, with the time-averaged fields satisfying the
conditions for a simple matter-dominated cosmology.  In this case $H$
can be approximated by a curve going as $2/3(t-t_0)$, describing a
dust model with a displaced origin.  The universe then appears as if
it has been generated by a big bang at a later time.  We call this the
effective big bang.  Of course, around this time we expect reheating
effects to start to dominate, so that in reality the universe must
pass over to a radiation-dominated era.  However, the naive `effective
big bang' concept is useful for extracting some qualitative
predictions from our model.

\begin{figure}
\begin{center}
\includegraphics[height=8cm,angle=-90]{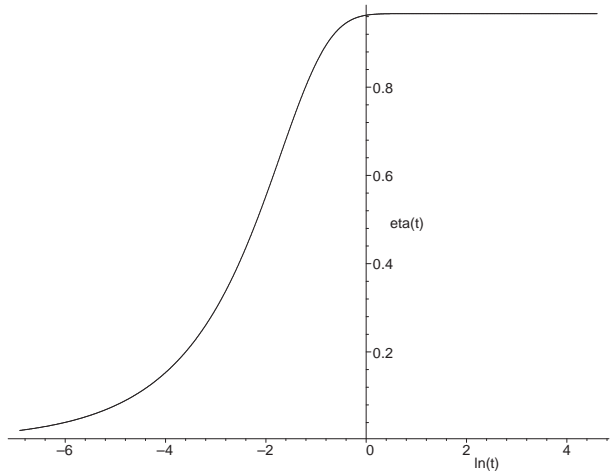}
\end{center}
\caption[dummy1]{\textit{Evolution of the conformal time $\eta(t)$ as
a function of $\ln(t)$.} The parameters for the model are as given in
figure~\ref{fHout} and again $t$ is measured in units of the Planck
time.  This plot sets $\mu=1$, so the time variable must be scaled to
correspond to more physical values of $\mu \approx 10^{-6}$.}
\label{fetaevol}
\end{figure}

As well as giving a new effective start time for the big bang,
inflation must also resolve the horizon problem.  The conformal
picture provides a straightforward means of visualising this process.
The key is to compute how much conformal time has elapsed by the end
of the inflationary era.  Using the same parameters as in
figure~\ref{fHout} we find that the conformal time evolves as shown in
figure~\ref{fetaevol}.  As the universe emerges from the (real) big
bang $\eta$ initially grows at $t^{2/3}$, as can be seen from
equation~\eqref{Rnear0}.  This gives rise to the rapid growth seen in
figure~\ref{fetaevol}, which plots $\eta$ as a function of
$\ln(t/t_p)$.  But once the inflationary region is entered, $R(t)$
starts to increase rapidly.  So the conformal time, which involves the
time integral of $R^{-1}$, quickly saturates, as can also be seen in
figure~\ref{fetaevol}.  Any further increase in $\eta$ will take an
extremely long time to occur, since the integrand $1/R$ is extremely
small.  So if $\eta$ has not reached $\pi/2$ before $R$ has inflated
significantly, the universe will have to exist for an extremely long
time to reach the boundary value of $\pi/2$.

\begin{figure}
\begin{center}
\includegraphics[height=7cm,angle=-90]{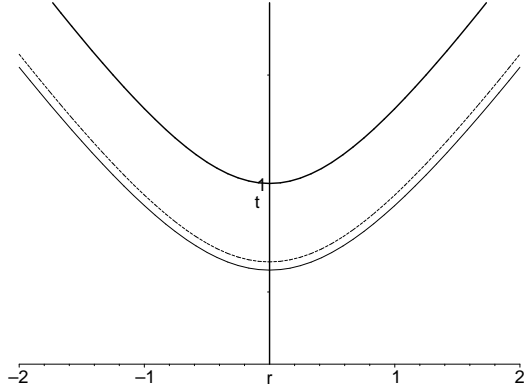}
\end{center}
\caption[dummy1]{\textit{The horizon problem for a closed universe in
    the conformal picture.} The lower solid line represents the surface
    of last scattering, and the broken line just above it is the
    current epoch.  The thick line is the future boundary
    ($t=\infty$).  Null geodesics lie at $45^\circ$, so there is no
    problem with points in the observable part of the surface of last
    scattering having once been in causal contact.}
\label{fhorprob}
\end{figure}

For the parameters used in figures~\ref{fHout} and~\ref{fetaevol} we
can see that $\eta$ saturates at a value of around 0.97 by the end of
inflation (at around $t=100 t_p$).  The value of $R$ at this time is
$2 \times 10^{22} \mu^{-1} l_p$.  All changes in $\eta$ beyond the end
of inflation occur very slowly.  Irrespective of the properties of the
reheating phase, $\eta$ will have changed very little by the time
recombination occurs.  This pushes the era of recombination to a much
later conformal time, close to the value for the present epoch.  This
removes the horizon problem, as the obervable patch of the universe
has been in causal contact and has had sufficient time to thermalise
(see figure~\ref{fhorprob}).  This is not true of the universe as a
whole, as antipodal points remain outside each other's light cone
until $t=\infty$.

In order to make some initial estimates from our boundary condition,
we will ignore the effects of reheating and nucleosynthesis, and
assume that the universe just smoothly enters a matter-dominated phase
described by
\begin{equation}
R = R_1 \left( \frac{t-t_0}{t_1-t_0} \right)^{2/3}.
\end{equation}
For the parameters we are currently considering we have $R_1 = 2
\times 10^{22} l_p$, $t_0 = 14.56 t_p$ and $t_1=100 t_p$.  We
will also ignore the cosmological constant, though this will clearly
affect the behaviour at late times.  The conformal time elapsed at
some later cosmic time $t$ can be approximated by
\begin{equation}
\eta \approx \eta(t_1) + \frac{3}{R_1} (t-t_0)^{1/3}
(t_1-t_0)^{2/3} \approx 0.97 + 2.4 \times 10^{-19} \, t^{1/3},
\end{equation}
where $t$ is measured in units of the Planck time.  We can therefore
arrive at a crude estimate for the age of the universe simply by
setting $\eta=\pi/2$.  This gives a time of $1.6 \times 10^{61}$ in
Planck units, or about $2.8 \times 10^{10}$ years.  This is a good
illustration of how some very large numbers (such as the age of the
universe in Planck times) can emerge from a simple model of inflation,
augmented with our closed universe boundary condition.

\begin{figure}
\begin{center}
\psfrag{Omm}{$\Omega_M$}
\psfrag{Oml}{$\Omega_\Lambda$}
\includegraphics[height=8cm,angle=-90]{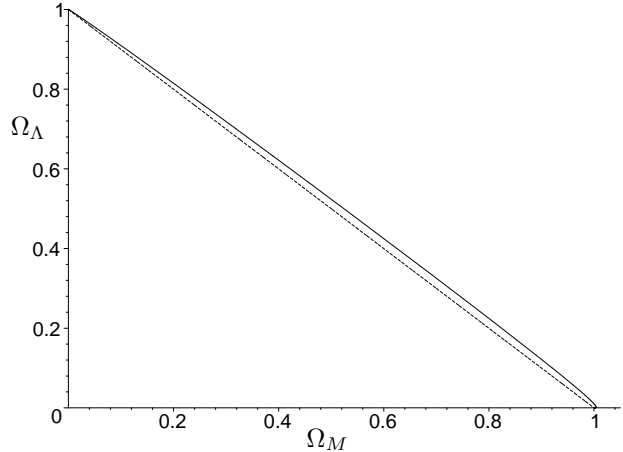}
\end{center}
\caption[dummy1]{\textit{Matter-dominated trajectory in the
$\Om_M$--$\Om_\Lambda$ plane.} The trajectory (solid line) is picked
out by demanding that the total conformal time elapsed throughout the
evolution is 0.6.  This evolution joins onto an inflationary model
which takes up around 0.97 units of conformal time, so that the
boundary condition of a total of $\pi/2$ is obeyed.}
\label{fMLeta}
\end{figure}

In order to produce a more detailed prediction, we return to
considering trajectories in the $\Om_M$--$\Om_\Lam$ plane.  We can
still assume that the universe is essentially matter dominated
throughout its history, but now the total amount of conformal time
that should elapse along the trajectory is less than $\pi/2$.  The
effect of this is to drive us onto trajectories which are closer to
the critical line.  If we only require a total conformal time of
around 0.6 to elapse during the matter-dominated phase of the universe
(as for the model above), then we require an $\alpha$ value of around
$10^4$, producing a trajectory that is very close to spatially flat,
and predicting a universe which is currently closed at around the
$1\%$ level (see figure~\ref{fMLeta}).

We can now see that our boundary condition gives rise to a remarkable
see-saw mechanism, with two key components.  The first component
relates to the amount of inflation that occurs.  For a range of
conditions, the time at which inflation starts can be approximated by
\begin{equation}
u_i = \frac{1}{(12 \pi)^{1/2} \mu b_0}.
\label{defui}
\end{equation}
A check on the physical approximations we make here is that this
quantity has the correct scale transformation properties under the
transformation of equation~\eqref{scl1}. The value of $\Hbar$ at the
start of slow-roll inflation can be approximated by $1/(3u_i)$,
\begin{equation}
\Hbar_i = \frac{1}{3u_i} ,
\end{equation}
and a suitable condition for the end of slow roll can be taken as
\begin{equation}
\Hbar_f^2 = \frac{\mu^2}{3}.
\label{Hend}
\end{equation}
During slow roll the evolution of $\Hbar$ is well approximated by a
linear fit, so
\begin{equation}
\Hbar = \Hbar_i - \frac{u-u_i}{u_f-u_i} (\Hbar_i - \Hbar_f).
\end{equation}
The gradient is determined from the field equations to be
\begin{equation}
\frac{\Hbar_i-\Hbar_f}{u_f-u_i} = \frac{\mu^2}{3}.
\end{equation}
The total number of e-foldings, $N$, can therefore be approximated by
\begin{equation}
N = \int_{u_i}^{u_f} \!\! \Hbar \, du \approx \frac{3}{2\mu^2}(\Hbar_i^2 -
\Hbar_f^2) \approx 2 \pi b_0^2.
\label{napprox}
\end{equation}
Numerical simulations reveal that this approximation does give a
correct order-of-magnitude for the number of e-foldings, though it
does tend to underestimate the precise value.  As expected, the result
for $N$ is invariant under scale changes, as is clear from the
integral formula for $N$.  The key conclusion is that the number of
e-foldings is primarily determined by $b_0$, with the value of $H$ at
the end of inflation fixed by $\mu$.  The value of $b_4$ essentially
decouples from $N$.  An increase in $b_0$ leads to an increase in the
final value of $R$, and hence to an increase in $\rho R^3$ in the
matter phase.  From equation~\eqref{defalp} we see that, in order to
preserve $\alpha$, any increase in $\rho R^3$ must be offset by a
decrease in the cosmological constant.  Conversely, the observed value
of $\Lambda$ can be employed to constrain the amount of inflation.

The second quantity of interest, given our boundary condition, is the
amount of conformal time that has elapsed before the universe enters
the present matter-dominated epoch.  Figure~\ref{fetaevol} shows the
typical evolution of $\eta$.  We see that a large percentage of the
conformal time is taken up before inflation starts.  During this
region we can approximate $R$ by equation~\eqref{Rnear0}.  Taking
$u_i$ as roughly approximating the point where $\eta$ starts to
saturate, we find that the conformal time that has elapsed is
approximated by
\begin{equation}
\eta = \int_0^{u_i t_p} \frac{dt}{R} \approx 0.92 \left(
\frac{|b_4|}{\mu^{4/3} }\right)^{1/2}
\left( \frac{1}{b_0^2}\right).
\label{etaapprox}
\end{equation}
And again, the result naturally assembles into a scale-invariant
quantity.  This result typically underestimates the total elapsed
$\eta$ at the end of inflation, but it does set a useful restriction
on the order of magnitude of the expression on the right-hand side.
We now see that, for fixed $b_0$ and $\mu$, the greater the initial
curvature (increasing $|b_4|$), the greater the amount of conformal
time taken up during inflation.  This in turn forces the universe
today to be closer to flat.  And conversely, the more spatially curved
we want the universe to be today, the closer to flat it must be
initially.  This is the second component of the see-saw mechanism.
The observed values of the cosmological parameters, together with our
boundary condition, therefore place a severe restriction on the
allowed models.

Given the strong constraints our model places on trajectories in the
$\Om_M-\Om_\Lam$ plane, the model must predict the small value of the
cosmological constant.  To see how small values of $\Lam$ naturally
arise, we return to equation~\eqref{defalp} and write
\begin{equation}
\Lam =
\frac{27 \alp}{(8 \pi G \rho_m R^3)^2}.
\end{equation}
In order to make some useful approximations we will again ignore the
effects of reheating and radiation domination, and simply assume that
$\rho_m R^3$ is given by its value for the scalar field at the end of
inflation.  If we furthermore ignore the effects of curvature and
$\Lam$ at this stage, we can write $8\pi G \rho_m = 3 H^2$, so we have
\begin{equation}
\Lam = \left. \frac{3 \alp}{H^4 R^6} \right|_{\mathrm{end}} .
\end{equation}
This expression does not depend on the precise moment of evaluation.
As we have seen, typical values of $\alp$ are in the range $10^{3}$
to $10^{5}$.  The small value of $\Lam$ is therefore seen to be a
consequence of the large values of $RH^{2/3}$ achieved by the end of
inflation.  The value of $H$ at the end of inflation is given by
equation~\eqref{Hend}, and for $R$ at the end of inflation, $R_f$, we can
write
\begin{equation}
R_f = e^N R_i.
\end{equation}
Here $R_i$ is the value of $R$ at the start of inflation, and $N$ is
the number of e-foldings.  We can estimate $R_i$ using
equation~\eqref{Rnear0} and the value of $u_i$ from
equation~\eqref{defui}.  Putting these together, we arrive at the
approximate formula
\begin{equation}
\Lam \approx 8 \times 10^4 \alp \frac{|b_4|}{\mu^{4/3}} b_0^2 \mu^2 e^{-6N}
l_p^{-2}.
\end{equation}
The typical models we consider have $N$ in the range of 40 to 50, and
$\mu$ around $10^{-6}$, which does indeed predict an extremely small
value for the cosmological constant.  As commented in the
introduction, the dominant behaviour in $\Lam$ is governed by the
$\exp(-6N)$ term, which immediately yields values for $\Lam$ of the
expected magnitude.

\section{The curvature spectrum}
\label{S-curv}

Our proposed `quantisation' condition on the available conformal time
places a strong restriction on the class of allowed models.  We must
now see whether any of these models produce a perturbation spectrum
consistent with the observed primordial fluctuations in the CMB.  Due
to the scale-invariant nature of the equations (under the
transformation of equation~\eqref{scltrf}), the parameters that
determine the key features of the inflationary period are the
scale-invariant quantities $b_0$ and $b_4/\mu^{4/3}$.  Furthermore,
the restriction that the total amount of conformal time available is
$\pi/2$ constrains the combination in equation~\eqref{etaapprox} to be
of order 1.  This limits us to a highly restricted class of models.

The scale transformation we have utilised at various points is not
conserved by quantisation, so vacuum fluctuations impose an absolute
scale on the model.  The key quantity that determines the size of the
perturbations is the mass ratio $\mu=m/m_p$.  For models with a
quadratic potential to generate perturbations consistent with
observation we need to set $\mu$ to be around $10^{-6}$.  As explained
above, this value is irrelevant for the evolution of the scalar field,
but $\mu$ does set the absolute size of the Hubble function, and so
fixes the magnitude of perturbations.

In order to consider a concrete example, we adopt the following set of
parameters:
\begin{equation}
b_0 = 2.37, \qquad \frac{b_4}{\mu^{4/3}} = - 0.49,
\qquad \mu = 2.1 \times 10^{-6}.
\label{paramvals}
\end{equation}
These values are typical for giving good agreement with experiment for
the curvature spectrum.  Evidently, little fine tuning is required in
chosing the two scale-invariant combinations, both of which are of
order unity.  Furthermore, the predictions we make here are quite
robust under variations of these parameters.  We are a long way
removed from areas of parameter space where chaotic dynamics of the
type considered by Cornish \& Shellard~\cite{cor98} could prove
significant.

\begin{figure}
\begin{center}
\includegraphics[height=9cm,angle=-90]{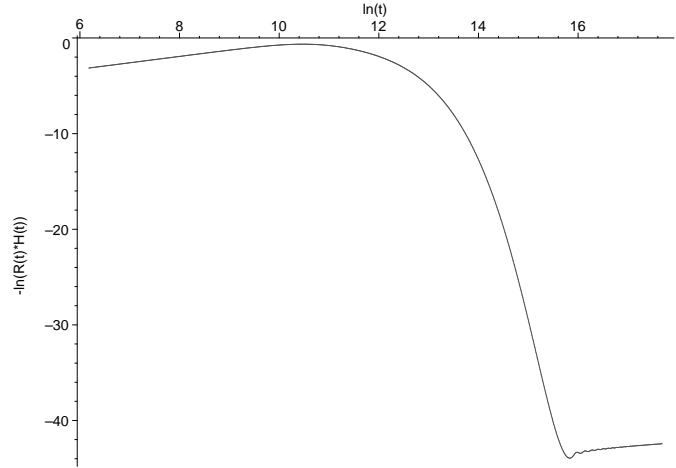}
\end{center}
\caption[dummy1]{\textit{Evolution of $1/(RH)$}.  The natural log of
 the dimensionless ratio of the Hubble radius to the curvature radius
 is plotted against log time (in units of $t_p$).  The parameters for
 the model are as described in the text.  During the inflationary
 period, $1/(RH)$ is a decreasing function of time, corresponding to
 an accelerating phase.}
\label{lnRH}
\end{figure}

In a closed universe model there are two scales of interest: the
horizon size $1/H$, and the radius of the spatial sections $R$.  The
ratio of these defines the dimensionless quantity $1/(RH)$, which is
related to the closure density by
\begin{equation}
\Om -1 = \frac{1}{(RH)^2},
\label{OmRH}
\end{equation}
where the total $\Om$ is defined by
\begin{equation}
\Om= \Om_\Lam + \Om_M.
\end{equation}
During inflation $\Om$ should \textit{decrease}, as the universe is
driven towards flatness.  This effect can be seen for our chosen
parameters in figure~\ref{lnRH}, which confirms that the universe is
accelerating over the range $e^{11} < t/t_p < e^{16}$.  The plot also
shows that the radius is always greater than the horizon size, which
is important when considering perturbation modes.

\begin{figure}
\begin{center}
\includegraphics[height=9cm,angle=-90]{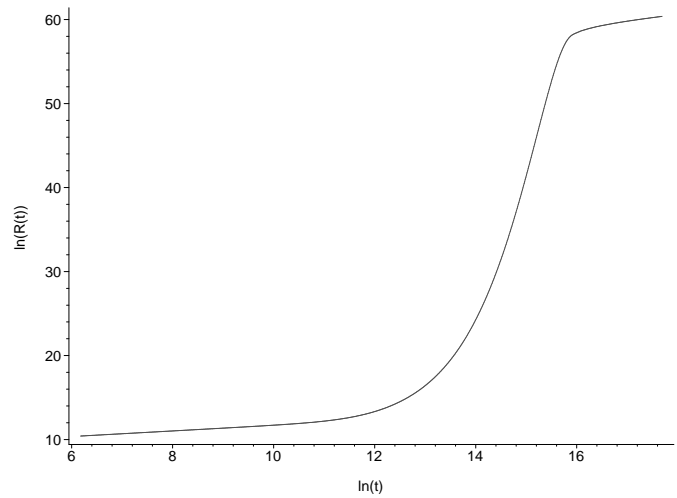}
\end{center}
\caption[dummy1]{\textit{Evolution of the scale factor}.  The
evolution of logarithm of the scale factor $R/l_p$ is shown as a
function of log time ($t$ in units of $t_p$).}
\label{sclfac}
\end{figure}

The evolution of the scale factor for our chosen parameters is shown
in figure~\ref{sclfac}.  During the inflationary region $\ln(R)$
increases from about 14 to 60.  This corresponds to a total number
of e-foldings of 46, as expected given our choice of parameters.  This
is roughly the figure we expect for the number of e-foldings if the
universe departs from flatness by an amount that is visible
today~\cite{sta96}.  In order to give rise to the observed fluctuation
spectrum it is thought that around 40--50 e-foldings are required
between the point where the present horizon scale leaves the horizon
and the end of inflation~\cite{LLinf}.  For this to hold, we therefore
require that the present horizon scale leaves the Hubble horizon soon
after the onset of inflation.

Closed universe models have an advantage over flat models when
discussing scales and perturbations, because equation~\eqref{OmRH}
gives a direct expression for $R$ in terms of $\Om$ and $H$.  To apply
this, suppose that at the current epoch a given physical size occupies
a fraction $x$ of the Hubble horizon $1/H_0$.  The corresponding
physical scale is
\begin{equation}
d_0 = \frac{x}{H_0}, \quad 0 < x < 1.
\end{equation}
For wave modes, $d_0$ will be quantised in units of $2\pi
R_0/(n^2-1)^{1/2}$, where $n$ is an integer.  We will ignore this
quantisation effect here, as the number of modes available ensure that
a continuum is a reasonable approximation for all but the low-$k$
modes.  Now consider the distance corresponding to $d_0$ in an epoch
when the scale factor is $R$ instead of $R_0$.  This scale is simply
$Rx/(R_0 H_0)$.  This distance is equal to the Hubble horizon at the
epoch corresponding to $R$ when
\begin{equation}
\frac{Rx}{R_0 H_0} = \frac{1}{H}.
\end{equation}
(As is standard practice we formulate the conditions for horizon
crossing in terms of the Hubble horizon, as opposed to the particle
horizon.)  It follows that the scale corresponding to a given $x$
crosses the Hubble horizon when
\begin{equation}
x = \frac{R_0 H_0}{RH} = \frac{\sqrt{\Om-1}}{\sqrt{\Om_0-1}}.
\label{xOm}
\end{equation}
So as $\Om$ is driven to unity during inflation, progressively smaller
scales cross the horizon as time advances.

\begin{figure}
\begin{center}
\includegraphics[height=10cm,angle=-90]{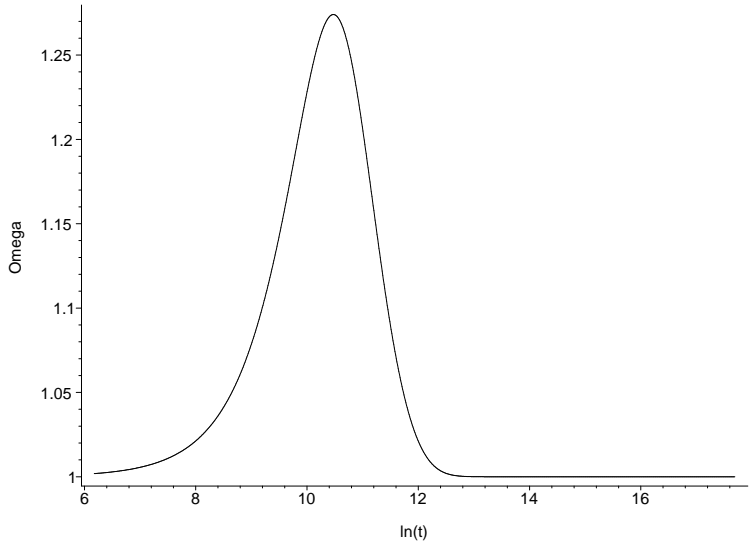}
\end{center}
\caption[dummy1]{\textit{Evolution of the total $\Om$ during
inflation}.  The total density is quickly driven close to flat, after
which $\Om$ only changes slowly.}
\label{Ominf}
\end{figure}

Figure~\ref{Ominf} shows the behaviour of $\Om$ in our scalar field
model.  As expected, $\Om$ is driven close to 1 in the region $e^{11}
< t/t_p < e^{16}$.  In order to use this plot in conjunction with
equation~\eqref{xOm} we need to assign a value to $\Om_0$.  The most
recent results from WMAP, with all data folded in, give a value of
$\Om_0=1.02 \pm 0.02$~\cite{WMAP}, so here we work with a value of
$\Om_0=1.02$.  The scale corresponding to the present horizon size
($x=1$) is therefore obtained during inflation when $\Om=1.02$.
Figure~\ref{Ominf} shows that this occurs at $\ln(t/t_p)= 12.0$.  By
reference to figure~\ref{sclfac}, we see that this time corresponds to
roughly 2 e-foldings.  This leaves roughly 44 e-foldings to the end of
inflation, which are sufficient to produce a fluctuation spectrum with
the desired properties.  These values are consistent with the
estimates of Uzan, Kirchner \& Ellis.~\cite{uza03}.

Each value of $x$ less than 1 corresponds to a scale which leaves the
Hubble horizon at successively smaller values of $\Om$.  A remarkable
feature of the approach outlined here is that it does not require any
knowledge of the evolution of the universe to relate scales today to
scales at the end of inflation.  This is because in any closed (or
non-flat) universe it is possible to measure $R$ directly.  This
cannot be done if the universe is assumed to be flat throughout
inflation, as one loses the ability to track scales through the
evolution of $\Om$.  In flat models the only way to find the scale
ratio $R/R_0$ is to evolve the scale factor $R$ through the history of
the universe, including the epoch of reheating, about which little is
known.  Of course, our method would be impossible to apply if the
value of $\Om_0$ turns out to be undetectably close to 1.  This is
predicted by approaches such as chaotic inflation, which can typically
allow of order $10^{10}$ e-foldings~\cite{LLinf}.  But for models that
allow $\Om_0$ to deviate measurably from 1 the method we have outlined
here should be of considerable use in calculating the fluctuation
spectrum induced by inflation.

In order to compute the primordial curvature spectrum we adopt the
slow-roll approximation, following the presentation of Liddle \&
Lyth~\cite{LLinf}.  For a quadratic potential, the slow roll condition
can be expressed simply in terms of the single parameter $\eps$, where
\begin{equation}
\eps = \frac{1}{4\pi \phi^2}.
\label{defeps}
\end{equation}
For slow roll to hold we require that $\eps \ll 1$.  However, the
model we consider here has $\eps$ in the range from $10^{-2}$ to
$10^{-1}$, with the larger values corresponding to higher $k$
values. We are therefore close to the point where slow roll
ceases to be a good approximation.  A concrete example of this is
seen when we consider the spectral index (see also Martin \&
Schwarz~\cite{mar00} for a discussion of the precision of slow
roll).  For the main evolution of $\phi$, however, plots such as
figure~\ref{fHfrom0} confirm that the slow roll approximations
are holding up, with the evolution of $H$ roughly linear with the
expected gradient of $\mu^2/3$.  A further source of error, as
stated earlier, is that we are ignoring the quantisation effects
implied by a closed cosmology. But, despite these provisos, we
may hope that the simple, slow-roll equations will still give a
reasonably accurate prediction for the curvature spectrum.

\begin{figure}
\begin{center}
\includegraphics[height=11cm,angle=-90]{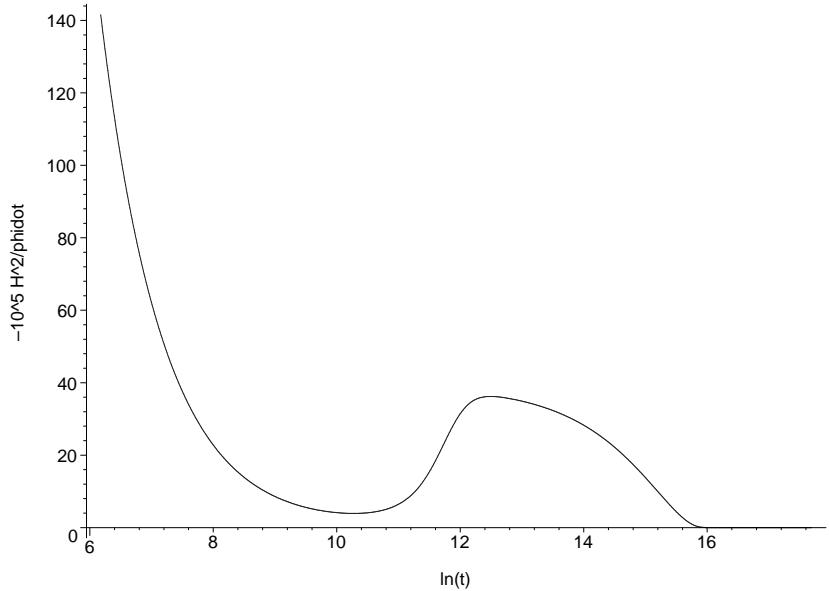}
\end{center}
\caption[dummy1]{\textit{Evolution of $H^2/\dot{\phi}$ during
inflation}.  The function plotted is $-10^5 H^2/\phidot$ as a function
of $\ln(t/t_p)$.  This function controls the magnitude of the curvature
perturbation.}
\label{Hphidt}
\end{figure}

Given the slow roll approximations, the key equation determining the
curva\-ture spectrum $\clp_\clr(k)$ is~\cite{LLinf}
\begin{equation}
\clp_\clr(k) = \left( \frac{H}{\dot{\phi}} \right)^2
\left( \frac{H}{2 \pi} \right)^2,
\label{PrK}
\end{equation}
where the right-hand side is evaluated when a given scale $x$ crosses
the horizon.  If we write
\begin{equation}
\clp_{\clr}^{1/2}(k) = \frac{-H^2}{2\pi\dot{\phi}}
\end{equation}
we see that the power spectrum is controlled by the ratio
$H^2/\dot{\phi}$.  This function is plotted in figure~\ref{Hphidt}.
Figures~\ref{Ominf} and~\ref{Hphidt} contain all of the information
required to extract $\clp_\clr(k)$.  Given a comoving wavenumber $k$,
we find $x$ from
\begin{equation}
\frac{1}{k} = \frac{x}{H_0}.
\end{equation}
The value of $x$ is converted to a time using equation~\eqref{xOm} and
the relationship plotted in figure~\ref{Ominf}.  With the value of $t$
then found, we can read off the spectrum from figure~\ref{Hphidt}.
The fact that the graph turns over for $\ln(t/t_p) < 12$ suggests that
the spectrum will contain less power at low-$k$ values than would be
expected from a straightforward power law in a flat cosmology, in
contrast to the predictions of Starobinsky~\cite{sta96}.  But again,
more detailed calculations are required to confirm this point.

\begin{figure}
\begin{center}
\includegraphics[height=10cm,angle=-90]{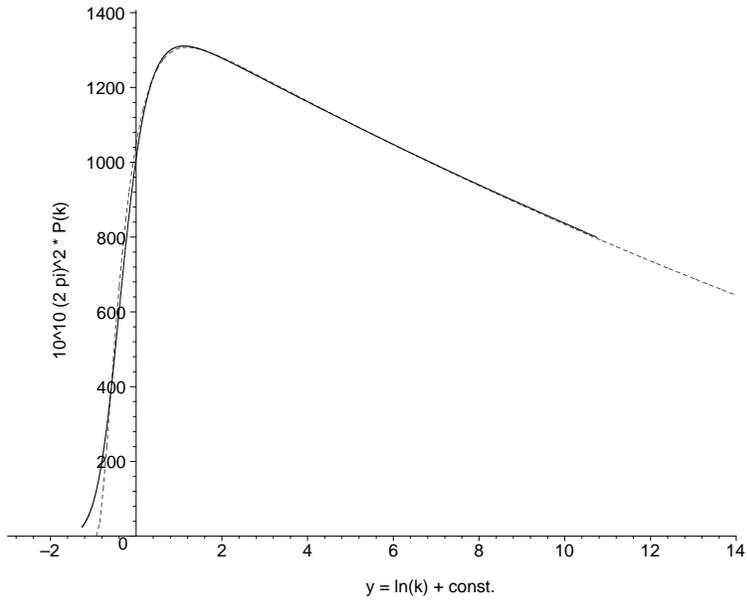}
\end{center}
\caption[dummy1]{\textit{The power spectrum of curvature
  perturbations}.  The function plotted is $10^{10} (2 \pi)^2
  \clp_\clr(k)$ as a function of $y=ln(1/x)=\ln(k)+
  \mathrm{constant}$.  The solid line represents the numerical
  predictions from our model, and the broken line is the best-fit
  power law with an exponential cutoff.  The parameters are described
  in the text.}
\label{sclpow}
\end{figure}

\begin{figure}[t!]
\begin{center}
\includegraphics[height=10cm,angle=-90]{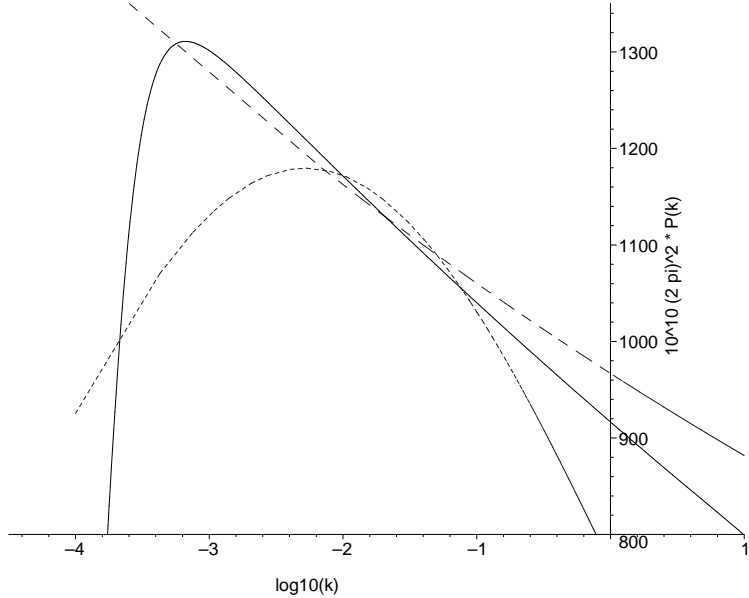}
\end{center}
\caption[dummy1]{\textit{Comparison of the scalar power spectrum with
    power law models.}  The function plotted is $10^{10} (2 \pi)^2
    \clp_\clr(k)$ as a function of $\log_{10}(k)$, assuming $h=0.65$.
    The solid line represents the numerical predictions from our
    model.  The long dashes represent the best fit power law
    ($n_s=0.96$) and the short dashes are the WMAP running spectral
    index best fit.  Notice that the vertical scale runs from 800 to
    1300, so the differences are slightly exaggerated.  }
\label{sclpow2}
\end{figure}

It is convenient to plot perturbation spectra as a function of
$y=\ln(1/x)$.  With dimensions reinserted, $y$ is given by
\begin{equation}
y = \ln \left( \frac{k}{h} \times 3 \times 10^3 \right)
\end{equation}
where $k$ is measured in $\mathrm{Mpc}^{-1}$ and the Hubble constant
is given by $H_0= 100 h \kmsmpc$.  In figure~\ref{sclpow} we plot the
power spectrum as a function of $y$ for our chosen parameters,
including $\Om_0=1.02$.  (For a plot showing the power spectrum as an
explicit function of $k$ in $\mathrm{Mpc}^{-1}$ see
figure~\ref{sclpow2}.)  The quantity plotted (solid curve) is $10^{10}
(2 \pi)^2 \clp_\clr(k)$. The graph shows a clear cutoff at low-$k$
values.  A cutoff of this type was proposed on phenomenological
grounds by Efstathiou~\cite{efs03}.  Indeed, it turns out that the
type of exponential cutoff proposed by Efstathiou agrees remarkably
well with our calculations, as can be seen in the dashed line, which
plots
\begin{equation}
10^{5} \, 2 \pi \, \mathcal{P}^{1/2}(k) = 37.6(1-.023
\,y)(1-\exp(-(y+0.93)/0.47)) ,
\label{spectral-fit}
\end{equation}
valid for $y>-0.93$.  The fact that our model predicts a spectrum
which has already been argued for on phenomenological grounds is very
reassuring.  We can now see that our model will produce a deficit in
the CMB power spectrum at low $\ell$, and in section~\ref{SCMB} we
confirm that this is the case.

Figure~\ref{sclpow} contains a further surprise.  From
equation~\eqref{spectral-fit} we see that at large $k$ the
relationship between $\clp_\clr^{1/2}(k)$ and $\ln(k)$ is roughly
linear, which is quite different from the expected power law relation.
To understand why this is the case we return to the slow roll
equations and derive the form of the spectrum for a quadratic
potential, following section~7.5 of Liddle and Lyth~\cite{LLinf}.  The
main relation we require relates the derivatives with respect to $k$
and $\phi$:
\begin{equation}
\frac{d}{d \ln k} = - \frac{1}{4 \pi \phi} \frac{d}{d\phi},
\end{equation}
where we specialise to results for a quadratic potential.  During slow
roll the following approximate relations hold:
\begin{equation}
H^2 = \frac{4 \pi}{3} \mu^2 \phi^2, \qquad \phidot =
\frac{\mu}{\sqrt{12 \pi}}.
\end{equation}
It follows that the curvature spectrum is approximated by
\begin{equation}
\clp_\clr^{1/2}(k) = 4 \left(\frac{\pi}{3}\right)^{1/2} \mu \phi^2.
\end{equation}
We therefore find that
\begin{equation}
\frac{d }{d \ln k} \clp_\clr^{1/2}(k) = - \frac{2 \mu}{\sqrt{3 \pi}}.
\end{equation}
This confirms that the relationship between $\clp_\clr^{1/2}(k)$ and
$\ln(k)$ is indeed linear.  Furthermore, the gradient is predicted to
be $-2\mu/\sqrt{3\pi}$.  For the model we are considering this
evaluates to $1.35 \times 10^{-6}$, which agrees well with the fit of
equation~\eqref{spectral-fit}, which has a gradient of $37.6 \times
0.023 /(2\pi \times 10^5) = 1.37 \times 10^{-6}$.  We can also see
that
\begin{equation}
\frac{d }{d \ln k} \ln(\clp_\clr(k)) = - 2
\frac{1}{\clp_\clr^{1/2}(k)} \frac{2 \mu}{\sqrt{3 \pi}} = -
\frac{1}{\pi \phi^2} = - 4 \epsilon
\end{equation}
where $\epsilon$ is the slow-roll parameter of
equation~\eqref{defeps}.  The typical slow-roll approximation is to
now write $\clp_\clr(k) \propto k^{-4 \epsilon}$, so that
\begin{equation}
\clp_\clr^{1/2}(k) \approx A k^{-2 \epsilon} \approx A (1 - 2\epsilon
\ln(k) + 2 \epsilon^2 \ln^2(k) + \cdots).
\end{equation}
The higher order terms here are incorrect, and introduce an error of
order $\epsilon \ln(k)$ to the power spectrum.  As commented earlier,
at high-$k$ values $\eps$ is in the range $0.01$--$0.1$, so this
deviation can become quite significant.

In figure \ref{sclpow2} we compare our computed scalar power spectrum
with the best fit power law ($n_s=0.96$) and the WMAP running spectral
index best fit.  The vertical normalisations for these fits have been
chosen arbitrarily, since it is the shape of the spectrum at large $k$
which is of greatest interest here.  The WMAP running
spectral index model shape is given by
\begin{equation}
\clp_\clr(k) = \clp_\clr(k_0) \exp\left((n_s-1)\ln(k/k_0)+\frac{1}{2}n_{\rm
run}(\ln(k/k_0))^2 \right).
\end{equation}
Using the WMAP figures for combined data sets their best fit values
are $n_s=0.93$, $n_{\rm run}=-0.031$ and
$k_0=0.05\impc$~\cite{WMAP}.  The running spectral index graph is
interesting in that it suggests that this model is attempting to
emulate both the cutoff at low $k$, as well as the reduced power at
large $k$.  The latter is possibly required as the strict power law
spectrum at large $k$ appears to over-predict the abundance of dwarf
galaxies.  This is discussed further in section~\ref{SCMB}.

A further quantity of interest is the tensor spectrum,
which is significant in searches for B-mode polarisation in the CMB.
As our definition of the tensor spectrum we take
\begin{equation}
\clp_{\mathrm{grav}}(k) = \frac{16}{\pi} H^2,
\label{tenspec}
\end{equation}
where again the right-hand side is evaluated as the scale controlled
by $x$ crosses the horizon.  (Here we follow the convention of Martin
\& Schwarz~\cite{mar00}).  The main quantity of interest is the ratio
of the tensor and scalar modes $r$, %
\begin{equation}
r = \frac{\clp_{\mathrm{grav}}}{\mathcal{P}_{\clr}},
\end{equation}
where the right-hand side is evaluated at some suitable low $k$.
Following the slow-roll approximation, we can write
\begin{equation}
\mathcal{P}_{\rm grav} (k)= \frac{64 \mu^2 \phi^2}{3}.
\end{equation}
Applying $d/(d\ln k)$ to this now yields
\begin{equation}
\frac{d}{d \ln k} \mathcal{P}_{\rm grav} (k)=
-\frac{32 \mu^2}{3 \pi},
\end{equation}
which is a constant.  For the tensor mode spectrum we therefore expect
that $\mathcal{P}_{\rm grav}$ should be linear in $\ln(k)$.

\begin{figure}[t!]
\begin{center}
\includegraphics[height=10cm,angle=-90]{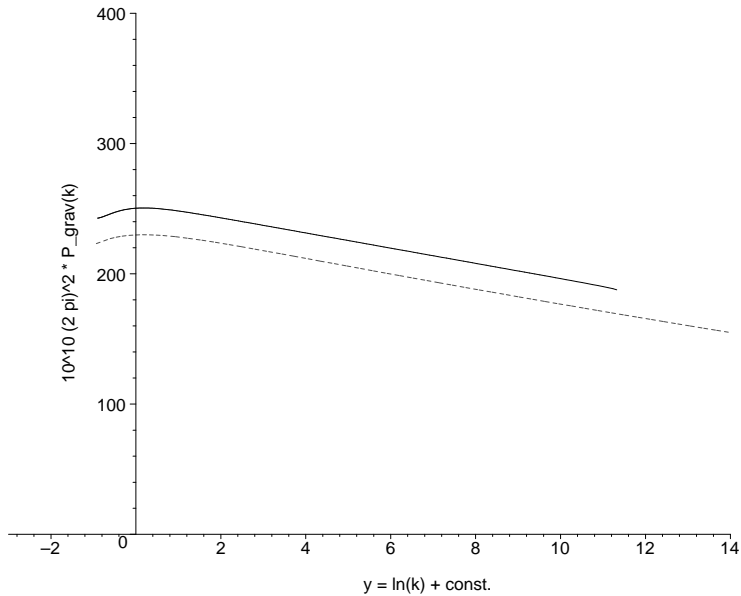}
\end{center}
\caption{\textit{Power spectrum of tensor perturbations}. The quantity
plotted is $10^{10} (2 \pi)^2 \mathcal{P}_{\rm grav}(k)$ versus
$y=\ln(k)+ \mathrm{constant}$.  The solid line is the numerical
predictions.  The dashed line is the best fit power law with
exponential cutoff, offset by 20 units vertically for clarity. All
parameters are described in the text.}
\label{Ftenspec}
\end{figure}

Figure~\ref{Ftenspec} shows the predicted tensor mode perturbation
spectrum for our model (again with $\Omega_0=1.02$).  Also shown
(slightly offset) is the best fit model of the form of
equation~\eqref{spectral-fit}, together with an exponential cutoff at
low $k$.  This best fit is given by
\begin{equation}
10^{10} \, (2 \pi)^2 \, \mathcal{P}_{\rm grav}(k) =
255(1-.023 \,y)(1-\exp(-(y+2.6)/0.65))
\label{tenspecfit}
\end{equation}
valid again for $y>-0.93$.  The slope in this formula is $255 \times
0.023 / (2 \pi \times 10^{5})^2 = 1.48 \times 10^{-11}$, which agrees
well with the theoretical value of $32 \mu^2/(3 \pi) = 1.47 \times
10^{-11}$.  If we form the ratio $r$ at $y=2$, which corresponds to $k
= 0.0016 \impc$ for $H_0 = 65 \kmsmpc$, we find $r \approx 0.19$. This
is large enough to be detectable in the near future.  For example, the
predicted sensitivity level to B modes for the Planck satellite
corresponds to a detection limit of $r>0.05$~\cite{fran03}, comfortably below
the value predicted here.  A relatively large tensor mode component is
therefore a firm prediction of our model.

We end this section with a discussion of how to overcome the
limitations of the slow-roll approximation and compute the
primoridal power spectrum more accurately from first principles.
The starting point for calculating the scalar curvature spectrum
in a simple flat model is to write the linearised perturbed
action in the form~\cite{muk92}
\begin{equation}
\del_2S = \frac{1}{2} \int d\eta \,d^3x \left( v'^2 - \eta^{ij}
v_{,i} v_{,j} + \frac{z''}{z} v^2 \right),
\end{equation}
where $v$ is a gauge-invariant combination of matter and metric
perturbations, and $z$ is defined by $z = \phi_0'/H$. Here
$\phi_0$ is the unperturbed field, and the dashes denote
derivative with respect to conformal time.  By working in this
way the entire problem is reduced to one of analysing a scalar
field with a time-dependent mass term.

This approach generalises to the non-flat case, although here the
definition of $z$ becomes more complicated~\cite{zhang03}. Also, for
closed models, the mode expansions necessary to carry out quantisation
have to be performed using spatial sections given by the 3-sphere
$S^3$.  In this case the `comoving wavenumber' $k$ takes on integer
values, with $k=3$ its lowest (non-gauge) value.  The mode equations
found this way are very complicated, but retain the general form
\begin{equation}
v_{k}'' + \left(k^2 - f(\eta,k)\right) v_{k} = 0.
\end{equation}
In the flat case $f(\eta,k)$ would be $z''/z$, and in all
cases $f(\eta,k)$ is calculable from knowledge of the background evolution.

The challenge now is to find suitable `quantum initial
conditions' so that after evolution through inflation, the
variables $v_k$ can be used to find the perturbation spectrum. To
achieve this involves a mode decomposition of $v_{k}$ into
positive and negative frequency modes.  The standard way to
approach this is to assume that in the asymptotic past the
background is either Minkowski or de Sitter, so that one knows
the correct vacuum to chose.  But this is clearly inappropriate
here, as looking back in time we encounter the initial
singularity.

The question of how to proceed in the absence of any asymptotic
notion of the vacuum state has been widely
discussed~\cite{bir-quant,par69,ful-qft}. One simple approach is
provided by Hamiltonian diagonalisation, where on each time slice
modes are decomposed into positive and negative energy states by
the Hamiltonian operator.  But this technique tends to
overestimate the particle production rate. A clear way to proceed
was developed by Parker and Fulling~\cite{par69,ful-qft} and
introduces the concept of an adiabatic vacuum. The application of
this to the present case is complicated, but possible
numerically, and initial results from this approach support a
low-$k$ cutoff, although with effects less pronounced than we
have found using the standard approximate techniques. This will
be discussed in future work.

\section{CMB power spectrum and large scale \\ structure}
\label{SCMB}

We now turn to using our primordial power spectra to compute the
measured fluctuations in the CMB.  For this purpose we use the CAMB
package~\cite{CAMB}, which takes as its input the curvature spectrum
and various cosmological parameters, and returns the $C_\ell$ values
of the CMB power spectrum.  For the cosmological parameters we take
$\Omega_{\rm cdm} h^2 = 0.135$, $\Omega_{\rm b} h^2 = 0.0224$ and the
optical depth to reionisation $\tau = 0.17$. Together with $\Omega_0 =
1.02$ these correspond to the WMAP best fit values~\cite{WMAP}.
Additionally, we use a Hubble constant of $65 \kmsmpc$.  This is lower
than the best fit value quoted WMAP of $72 \pm 5 \kmsmpc$.  This
latter value is obtained for an assumed {\em flat\/} model. If one
relaxes the assumption of flatness, then the CMB data is quite
degenerate with respect to the values of $H_0$ and $\Omega_{0}$ (see
section~6.2 of Spergel et al.~\cite{WMAP} for a discussion of this
point).  Our value of $H_0= 65 \kmsmpc$ is compatible at $1\sigma$
with virtually all current determinations of $H_0$.  For example, the
HST Key Project value is $72 \pm 8 \kmsmpc$~\cite{free01}, whereas
combined Sunyeav--Zel'dovich and X-ray flux measurements consistently
favour a slightly lower value~\cite{jones01}.  A lower value of $H_0$
was also suggested for closed models by Efstathiou~\cite{efs03}, and
our chosen value provides a reasonable fit with the WMAP power
spectrum for $\Omega_0=1.02$.  It is worth pointing out in this
context that the WMAP simultaneous best fit values of $H_0$ and
$\Om_0$ actually provide a rather poor fit to their measured CMB
spectrum!

\begin{figure}
\begin{center}
\includegraphics[height=10cm,angle=-90]{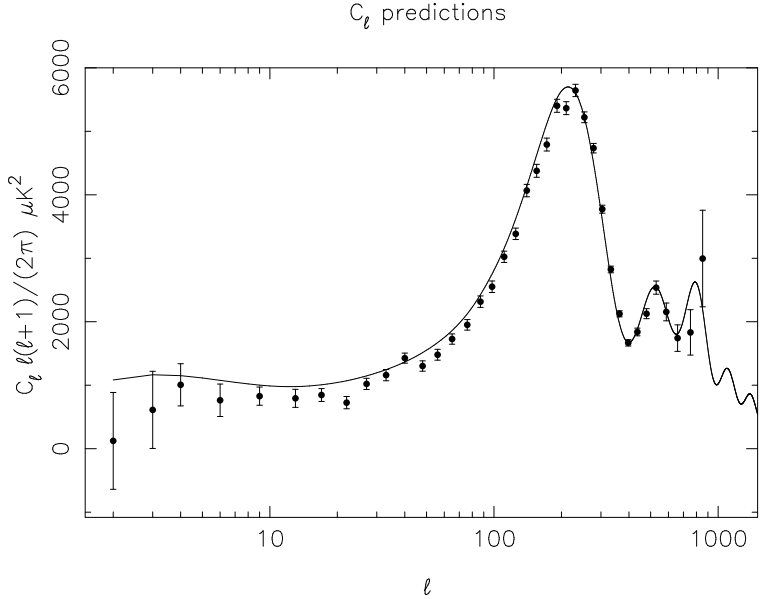}
\end{center}
\caption{\textit{Predicted CMB power spectrum}.  These predictions are
  for a model with $\Omega_{0} = 1.02$, as discussed in the
  text.  The experimental points shown with $1\sigma$ error bars are
  the WMAP results. }
\label{fCMB}
\end{figure}

In fixing $\Om_M$ and $\Om_0$ we in turn imply a value of $\Om_\Lam$.
Together these pick out a unique path in the $\Om_M$--$\Om_\Lam$
plane, for which we can calculate a total elapsed conformal time.  In
order to satisfy our boundary condition we need to achieve a value of
conformal time by the end of inflation of $0.975$.  The value of
$b_4/\mu^{4/3}$ introduced in equation~\eqref{paramvals} was chosen to
achieve this.  With all the parameters now chosen, the computed CMB
power spectrum is shown in figure~\ref{fCMB}.  The fit is reasonably
good, particularly at low $\ell$ where our model does show a fall off
in power.

\begin{figure}
\begin{center}
\includegraphics[height=10cm,angle=-90]{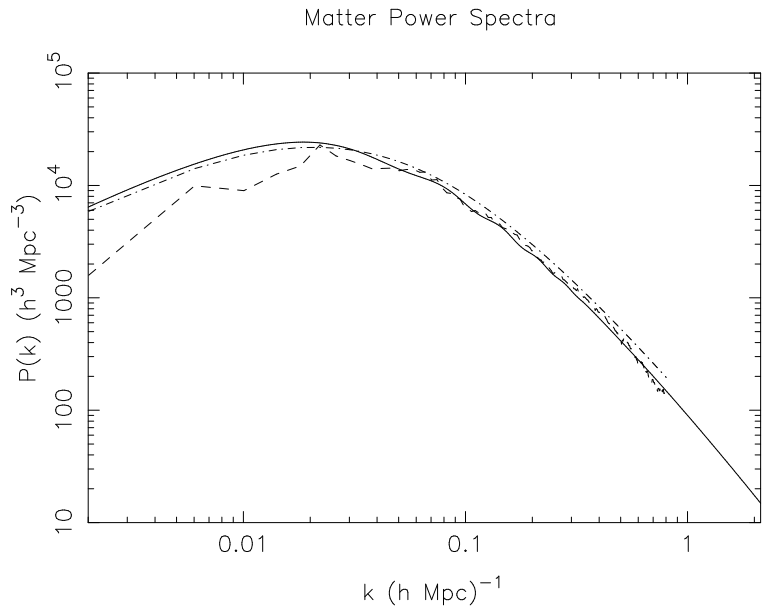}
\end{center}
\caption{\textit{The linear matter power spectrum}.  The solid
line is the prediction from CAMB using as input our
scalar primordial power spectrum. Long dashes are the
2dF measurements~\cite{perc01} convolved with the survey window function.
The dot-dash line is a comparison fiducial model with $n_s=1$,
$\sigma_8=1$, $h=0.7$, $\Omm h=0.2$, and $\Omega_b=0$.}
\label{fmatt}
\end{figure}

A further test of our model is the predicted linear matter power
spectrum.  This is shown in figure~\ref{fmatt}, where it is compared
with the 2dF data of Percival et al.~\cite{perc01}.  Also plotted is
a fiducial model with $n_s=1$, $\sigma_8=1$, $h=0.7$, $\Omm h=0.2$,
and $\Omega_b=0$ (and therefore no baryon oscillations).  This model
is calculated from the fitting formulae of Eisenstein \&
Hu~\cite{eis98}.  In order to achieve a good fit with the data, the
vertical scale for our model predictions has been adjusted downwards
by about 19\%.  Clearly, the model we have used so far needs some
adjustment in its chosen parameters to correct the normalisation.  But
here we only we wish to compare the overall shapes of the three matter
power spectra, leaving the overall normalisation aside.  It is clear
that the spectrum based on our non-power law primordial spectrum fits
the 2dF data with a similar degree of fidelity to the power law-based
fiducial model. On large scales (small $k$) both appear to
overestimate the 2dF results.  This can be explained by the fact that
the 2dF results are given convolved with the survey window function,
which has its largest effects at large scales (see figure~2 of
Percival et al.).  On smaller scales our results fit the shape of the
2dF spectrum very well.  There is therefore no apparent problem in
having a non-power law primordial spectrum right down to these
scales. On even smaller scales the relevant comparison would be with
Lyman-$\alpha$ data, but this area is still controversial and will not
be discussed here.

\begin{figure}
\begin{center}
\includegraphics[height=10cm,angle=-90]{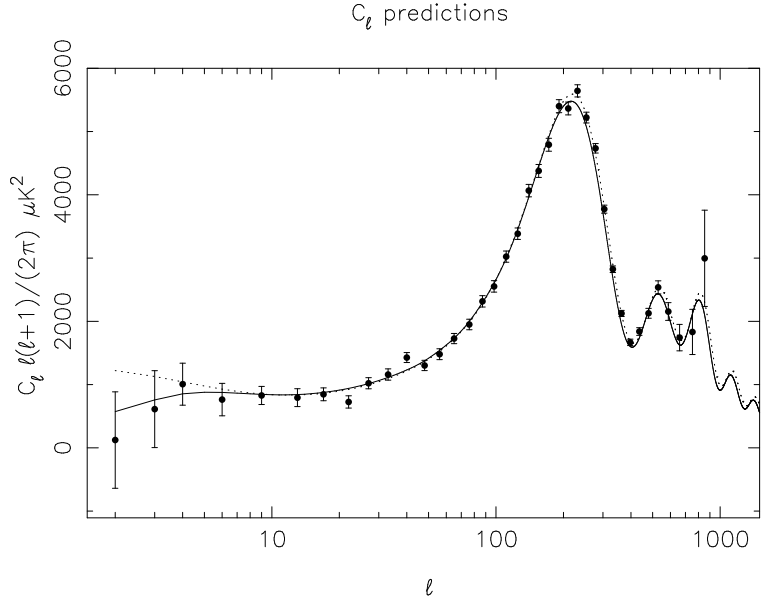}
\end{center}
\caption{\textit{CMB power spectrum for a model
with $\Omega_0 = 1.04$} (I).  The parameters are discussed in the text.
The experimental points shown with $1\sigma$ error
bars are the WMAP results~\cite{WMAP} and the dashed curve
corresponds to the best fit $\Lambda$ CDM power law
CMB power spectrum as distributed in the WMAP data
products.}
\label{fp04}
\end{figure}

\begin{figure}
\begin{center}
\includegraphics[height=10cm,angle=-90]{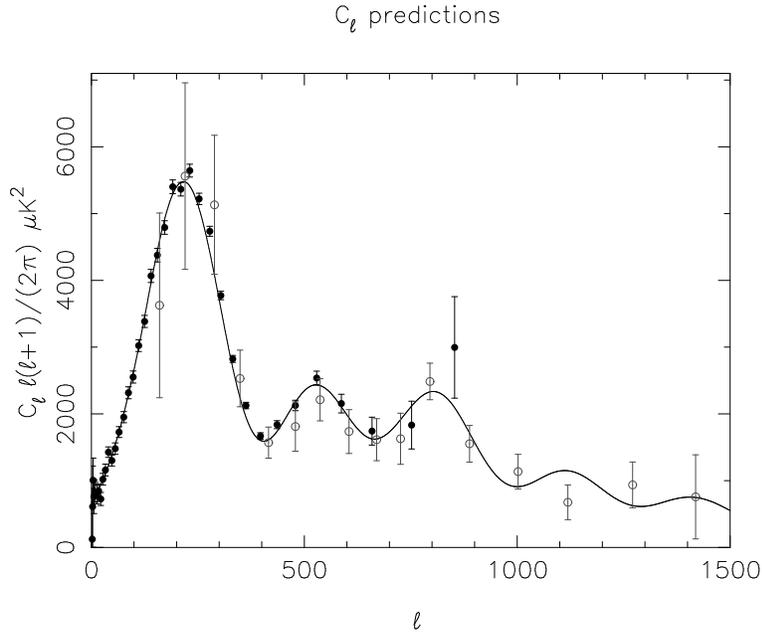}
\end{center}
\caption{\textit{CMB power spectrum for a model with $\Omega_0 = 1.04$}
(II). A similar plot to figure~\ref{fp04}, but now plotted on a
linear scale, with the latest Very Small Array data points also
included to extend out to higher $l$ values (open circles).}
\label{fvsa}
\end{figure}

So far we have developed a model based on the best fit WMAP values for
$\Omega_{\rm cdm}h^2$, $\Omega_{\rm b}h^2$ and $\Om_0=1.02$, but using a
different value of $H_0$.  This model does provide a reasonable fit,
but figure~\ref{fCMB} clearly leaves room for improvement, as does the
normalisation of the matter spectrum.  Ideally we would employ an MCMC
analysis to find the best fit parameters for our model, but here we
simply give an alternative model which improves the fit.  This model
has $\Om_0=1.04$, so it still consistent with WMAP at the $1\sigma$
level.  We also take $\Omega_{\rm b}h^2=0.0224$ (as before), $h=60$
and $\Omega_{\rm cdm} h^2 = 0.110$.  Together these yield a value of
$\Omega_{\rm cdm}=0.306$, which is reasonable.  The values of $b_0$,
$b_4$ and $\mu$ consistent with these are
\begin{equation}
b_0=2.3310, \qquad \frac{b_4}{\mu^{4/3}} = -0.2798, \qquad \mu = 1.94
\times 10^{-6}.
\end{equation}
The lower value of $|b_4|$ produces a universe with larger curvature
today.  The results of this model are summarised in a best-fit
curvature power spectrum given by
\begin{equation}
10^{5} \, 2 \pi \, \mathcal{P}_{\clr}^{1/2}(k) = 35.02(1-.023
\,y)(1-\exp(-(y+0.20)/0.397)).
\label{fit-1p04}
\end{equation}
The result for the CMB power spectrum with these parameters is shown
in figures~\ref{fp04} and~\ref{fvsa}.  The former includes the WMAP
points, and the latter plots both WMAP and the latest Very Small Array
(VSA) data points, plotted on a linear scale~\cite{vsa04}.  This model clearly
gives an extremely good fit to the observed spectrum, with the dip at
low $\ell$ now quite pronounced.  This is because an increase in
$\Om_0$ has the effect of sliding the graph to the right in $k$ space,
as is clear from equation~\eqref{xOm}.  There is also a good fit with
the VSA data at higher $l$ values.  The main disadvantage with this
model is the low value of $H_0$, though this is still within
$1.5\sigma$ of the HST Key Project value.


\section{Discussion}
\label{Sdisc}

Many authors have argued that it is difficult to obtain realistic
models of a closed inflationary
universe~\cite{lin03,sta96,uza03,ellis02a,ellis02}.  One of the main
problems is the low number of e-foldings required, which it is argued
demand an unacceptable degree of fine tuning in the initial conditions
for inflation.  We have shown that this is not the case.  It is
relatively straightforward to construct inflationary models resulting
in a universe which is measurably closed today, using only the
simplest model of a free massive scalar field.  These models have a
number of attractive features and provide a very good fit to the
observational data.

Motivated by the global geometry of de Sitter space, our model
incorporates a quantisation condition on the available conformal time.
This may seem uncomfortable from the point of view of causality, but
it is quite sound from the viewpoint of wave mechanics.  In field
theory one regularly applies boundary conditions at temporal infinity,
and the model presented here extends this idea to the universe.
Furthermore, the constraint does not relate to the actual conformal
time elapsed in the universe, but to the available conformal time at
the point of creation of a homogeneous, idealised universe, before
substantial structures have formed.

We hope that this quantisation condition will serve as a guide in
constructing more detailed physical models of the processes taking
place around the big bang, and that it might emerge from a quantum
gravity understanding of this phase.  Of course, many quantum gravity
models of the very early universe involve tunneling from a Euclidean
phase, in which case one cannot argue against the constraint on
grounds of causality.  Furthermore, for those who remain unhappy with
our proposed constraint, we would point out that many of our
conclusions do not depend crucially on it.  One can relax the
condition, explore the full space determined by the input parameters,
and still make sensible predictions about the primordial power
spectrum.

With a massive scalar field included, we find that initial data can be
specified in terms of a pair of scale-invariant coefficients in the
expansion of the scalar field around the initial singularity.  These
then determine the initial conditions for inflation.  Quantum gravity
effects will not be significant until well before the onset of
inflation, so we are fully justified in running the evolution
equations back to much closer to the initial singularity.  Given only
that we require of the order of 44 e-foldings we can conclude that the
scale-invariant parameters $b_0$ and $b_4/\mu^{4/3}$ are both of order
unity.  Because we start close to the initial singularity, with
infinite $H$, the model \textit{predicts} a period of inflation in the
early universe, independent of any question of generating
super-horizon fluctuations.  Furthermore, our condition on the total
conformal time ensures that the matter-dominated plot in
figure~\ref{fig8} represents the \textit{largest} departure from flat
that we can consider.  The inflationary period drives us closer to
flat and, assuming $\Om_M\approx 0.3$ we are led to an $\Om_\Lam$ of
around 0.7.  Viewed in reverse, given input parameters $b_0$ and
$b_4/\mu^{4/3}$ of order unity, together with our boundary condition,
we predict an extremely small value for $\Lam$, consistent with
observation.  The final parameter in the model is the mass ratio $\mu
= m/m_p$, which is fixed by considering the growth of perturbations
during the inflationary period.

The predictions of our model are robust against small variations of
the input parameters.  There is no question of the initial data being
specified in a chaotic region of the classical dynamics.  Chaotic
behaviour in the inflationary equations in a closed universe was
noticed by Page~\cite{page84} and Cornish \& Shellard~\cite{cor98}.
But these authors considered initial data around $H=0$, whereas our
parameters specify the fields as they exit the initial singularity,
with $H=\infty$.  The evolution therefore never enters the region of
parameter space where chaotic evolution becomes relevant.  The main
element of fine tuning required in our model is setting $\mu = m/m_p$
to $\approx 10^{-6}$, but this type of tuning of the potential is a
well-known feature of all inflationary theories~\cite{brand01}.
Masses of the order of $10^{-6}m_p$ are quite legitimate from the
viewpoint of high-energy physics, which we expect will shed light on
the possible origins of such particles.

Our model makes a series of predictions.  The universe is closed at
around the level of a few percent, and we can explain the dip in the
CMB power spectrum observed at low $\ell$.  The statistical
significance of this dip has been discussed by many
authors~\cite{efs03,efs03a,efs04,cont03}.  Their main conclusion is
that this dip is not inconsistent with the spatially flat concordance
model, but could still be an indicator of new physics, provided that
any new model does not introduce too many additional parameters.  In
this respect our model is very efficient.  Taking $\Om_M$ and
$\Om_\Lam$ as input data introduces two constraints on the three
parameters $b_0$, $b_4$ and $\mu$.  The one remaining degree of
freedom is removed by fixing the scale of the overall perturbation
amplitude.  We also predict a running spectral index with reduced
power at high-$\ell$ values.  This looks to be favoured by dwarf
galaxy and Lyman $\alpha$ data, though this area is controversial.
The behaviour of the CMB spectrum beyond the second peak will also
help break degeneracies in various models, and high-resolution
measurements in this region would be extremely valuable.

For our best-fit models we typically predict of the order of 50
e-foldings.  Such values are quite natural in our scheme, removing the
central objection to closed universe cosmologies.  Furthermore, a
value of 50 has the additional feature of removing any potential
problems with trans-Planckian physics~\cite{brand01}.  That is, there
is no question of Planck scale physics being inflated up to
cosmological scales.  A further indicator that our predictions are
relatively insensitive to quantum corrections is provided by
equation~\eqref{Etot0}, which gives the value of the total energy in
the universe as we leave the big bang.  Assuming that $b_4/\mu^{4/3}$
is of order unity, this expression becomes
\begin{equation}
E_{\mbox{tot}}  \approx \frac{0.03}{\mu^2}  \frac{\hbar}{t}.
\end{equation}
This tell us that the action $Et$ is very large, when measured in
units of $\hbar$.  This is reassuring, as it justifies treating the
universe on the whole as a classical object.  While quantum gravity
effects will certainly alter how the universe behaves around the
singular region, there is good reason to believe that the predictions
of this paper will be largely unaffected by any such new physics.
Furthermore, the limiting value of the action as $t \mapsto 0$ is
large in units of $\hbar$, suggesting that the universe started in a
highly improbable state.  This will naturally favour
spherically-symmetric initial conditions, and could also prove
attractive from the viewpoint of the extremely low entropy that the
universe has to have initially.

One further aspect of our model where considerations involving energy
are relevant is the end of inflation, as the universe enters a
reheating phase.  Converting the energy density at the end of
inflation into a temperature yields the reasonable value of $10^{16}$
GeV.  This is as expected for typical reheating, if it occurs
immediately after inflation is over.  But as the scalar field exits
inflation the universe is in a matter-dominated phase.  Some process
has to intervene which converts the scalar field into a
radiation-dominated phase.  Curiously, this may be difficult to
achieve in a way which preserves the total, integrated energy, though
there is little difficulty in demanding local conservation.  This is
probably a reminder that the `total energy' in a closed universe
remains a problematic concept in general relativity.

An important question is to identify the main experimental
observations that will argue for or against our model.  Our first key
result is the low-$\ell$ dip in the power spectrum.  The significance
of this could be further enhanced by any lowering of the error bars in
these measurements.  While cosmic variance is the ultimate determining
factor there is still some room for improvement, particularly in the
question of foreground contamination~\cite{efs03a}.  The drop in the
power spectrum at large $\ell$ is also a characteristic feature, which
could be detectable with future observations.  The third key
prediction is the high value of $r$ --- the ratio of tensor to scalar
perturbations.  This should also be observable by experiments such as
the Planck satellite.  It would also be of great interest to see which
of these predictions is dependent on our choice of potential, and
which survive if this restriction is removed.

We hope that much of the future work suggested by this paper is self
evident.  Before significant future work can begin we need to address
the two main provisos mentioned in the text.  The first is that mode
quantisation in a closed universe has to be built in.  The second is
that the mode equations need to be integrated numerically, as aspects
of the slow roll approximation are inadequate for our model.  Neither
of these pieces of work should prove too difficult, and a start has
been made by Zhang and Sun~\cite{zhang03} who appear to support our
use of equation~\eqref{PrK} as a leading order approximation.  One
area that we aim to address quickly is to explore in more detail the
effects of varying the main input parameters, to explore the range of
possible universes expected today.  One significant point here is that
all evolution is constrained to lie inside the matter curve of
figure~\ref{fig8}.  This in itself is a significant restriction.  In
addition, the parameters chosen for the examples presented here were
not computed by detailed model fitting.  A full Bayesian determination
of these parameters using all available data is clearly desirable, and
will give a more stringent test of how well our model holds up.

\section*{Acknowledgements}

We thank Anthony Challinor, Anze Slosar, Michael Hobson and other
members of the Cambridge--Leverhulme collaboration for useful
discussions.  CD Thanks the EPSRC for their support.

\begin{appendix}

\section{Conformal line elements}
\label{appa}

Here we establish spacetime conformal forms for the standard
cosmological line elements for open, closed and flat cosmologies. (The
fact that the Weyl tensor vanishes for cosmological models ensures
that this is possible.)  Our starting point is the FRW line element in
the form
\begin{equation}
ds^2 = dt^2 - \frac{R(t)^2}{(1+k r^2/4)^2} \bigl(dr^2 + r^2 (d\theta^2
+ \sin^2(\theta) \, d\phi^2) \bigr).
\end{equation}
This line element is already conformal in its spatial component. But a
problem with this form is that $r$ is assumed to be dimensionless.  To
rectify this we introduce a constant distance $\lambda$ and replace
the line element with
\begin{equation}
ds^2 = dt^2 - \frac{4 \lam^2 R(t)^2}{(\lam^2 + k r^2)^2} \bigl(dr^2 +
r^2 (d\theta^2 + \sin^2(\theta) \, d\phi^2) \bigr).
\end{equation}
Here $t$, $r$, $\lam$ and $R$ all have units of distance (assuming
$c=1$). As usual, the constant $k$ is either $\pm 1$ or zero.

We seek a coordinate transformation to the spacetime-conformal line
element
\begin{equation}
ds^2 = \frac{1}{f^2} \bigl( d \cdt^2 - d\cdr^2 - \cdr^2 (d\theta^2
+ \sin^2(\theta) \, d\phi^2) \bigr) ,
\end{equation}
where $f$ is a (dimensionless) function of $\cdt$ and $\cdr$.
Clearly, from the angular terms, we must have
\begin{equation}
\frac{\cdr}{f} = \frac{2 \lam r R(t)}{\lam^2 +k r^2}.
\end{equation}
In addition, the coordinate transformation must satisfy
\begin{align}
d\cdt &= f \cosh(u) \, dt + \frac{\cdr}{r} \sinh(u) \, dr \nn \\
d\cdr &= f \sinh(u) \, dt + \frac{\cdr}{r} \cosh(u) \, dr .
\end{align}
For flat cosmologies ($k=0$) we simply set $\cdt/\lam$ equal to the
conformal time $\eta$, with $\cdr=r$ and the hyperbolic angle $u$ set
to zero.  For non-flat cosmologies it is perhaps surprising to find
that $u$ is non-zero.  That is, there is a mismatch between the
conformal coordinate frame and the cosmological frame.

The integrability conditions for the coordinate transformation provide
a pair of differential equations for $u$:
\begin{align}
r \deriv{u}{r} &= \sinh(u) \label{rderiv1} \\
2 \lam r R(t) \deriv{u}{t} &= (\lam^2 + k r^2) \cosh (u) +
( - \lam^2 + k r^2) \label{rderiv2}
\end{align}
Equation~\eqref{rderiv1} can be solved straightforwardly to give
\begin{equation}
\et{u} = \frac{1 + r A(t)}{1-r A(t)}.
\end{equation}
The function $A(t)$ is determined by equation~\eqref{rderiv2}, which
reduces to
\begin{equation}
2 \lam R \dot{A} = \lam^2 A^2 + k,
\end{equation}
where the overdot denotes the derivative with respect to cosmic time
$t$.  The solution for $A$ depends on the curvature:
\begin{equation}
\lam A =
\begin{cases}
\tan(\eta/2) & k=1 \\
-\tanh(\eta/2) & k=-1.
\end{cases}
\end{equation}
The constant of integration has been chosen such that $A=0$
corresponds to $\eta=t=0$.

Next we solve the following pair of equations for $\cdr$:
\begin{equation}
r \deriv{\cdr}{r} = \cdr \cosh(u), \qquad
\deriv{\cdr}{t} = f \sinh(u).
\end{equation}
With a suitable choice of the arbitrary scale factor we find that
\begin{equation}
\cdr = \frac{\lam^2 A^2/k + 1}{1-r^2 A^2} \,  r.
\end{equation}
The remaining equations determining $\cdt$ are
\begin{equation}
r \deriv{\cdt}{r} = \cdr \sinh(u), \qquad
\deriv{\cdt}{t} = f \cosh(u).
\end{equation}
These are solved by
\begin{equation}
\cdt = \frac{\lam^2 +k r^2}{1-r^2 A^2} \, \frac{A}{k}.
\end{equation}
Here the arbitrary constant of integration has been fixed to ensure
that $t=\eta=0$ corresponds to $\cdt=0$.

With $\cdr$ and $\cdt$ now found, we can write
\begin{equation}
f = \frac{\lam^2 A^2 + k}{2 \lam AR} \cdt.
\end{equation}
Our remaining task is to find an expression for $t$ in terms of
$\cdt$ and $\cdr$.  For this we use that result that
\begin{equation}
\frac{2 \lam \cdt}{\lam^2 + k(\cdr^2 - \cdt^2)} = \frac{2 \lam A}{k -
 \lam^2 A^2}.
\end{equation}
The right-hand side is now a function of $t$ only, so the left-hand
side can be used in place of cosmic time.  Again, the explicit form
depends on the cosmology.  For spatially closed cosmologies we can write
\begin{equation}
\frac{2 \lam \cdt}{\lam^2 + \cdr^2 - \cdt^2} = \tan(\eta)
\end{equation}
whereas for spatially open we have
\begin{equation}
\frac{2 \lam \cdt}{\lam^2 + \cdt^2 - \cdr^2} = \tanh(\eta).
\end{equation}
The three cosmological scenarios can now be summarised as
\begin{align}
f &= \frac{\cdt}{R \sin(\eta)} =
g \left( \frac{2 \lam \cdt}{\lam^2 + \cdr^2 - \cdt^2} \right) \,
\frac{\cdt}{\lam} &
\mbox{closed} \nn \\
f &= \frac{2 \lam}{R} & \mbox{flat}  \\
f &=\frac{\cdt}{R \sinh(\eta)} = \bar{g} \left(\frac{2 \lam
  \cdt}{\lam^2 + \cdt^2 - \cdr^2} \right)  \,
\frac{\cdt}{\lam} & \mbox{open} \nn
\end{align}
which are the results used in the main text.

\end{appendix}


\begin{thebibliography}{10}

\bibitem{WMAP}
{D.N. Spergel et al}.
\newblock First year {W}ilkinson anisotropy probe ({WMAP}) observations:
  Determination of cosmological parameters.
\newblock {\em Astrophys. J. Suppl.}, 148:175, 2003.

\bibitem{brannan-cup}
D.A. Brannan, M.F. Espleen, and J.J. Gray.
\newblock {\em Geometry}.
\newblock Cambridge University Press, 1999.

\bibitem{gap}
C.J.L Doran and A.N. Lasenby.
\newblock {\em Geometric Algebra for Physicists}.
\newblock Cambridge University Press, 2003.

\bibitem{lin03}
A.~Linde.
\newblock Can we have inflation with {$\Omega > 1$}?
\newblock {\em J. Cosmol. Astropart. Phys.}, 5:2, 2003.

\bibitem{sta96}
A.A. Starobinsky.
\newblock Spectrum of initial perturbations in open and closed universes.
\newblock In M.~Khlopov, M.E. Prokhorov, A.A. Starobinsky, and J.~{Tran Thanh
  Van}, editors, {\em Cosmoparticle Physics}, page~43. Edition Frontiers, 1996.

\bibitem{uza03}
J.~Uzan, U.~Kirchner, and G.F.R. Ellis.
\newblock {WMAP} data and the curvature of space.
\newblock {\em MNRAS}, 344:L65, 2003.

\bibitem{ellis02a}
G.F.R. Ellis, W.~Stoeger, P.~McEwan, and P.~Dunsby.
\newblock Dynamics of inflationary universes with positive spatial curvature.
\newblock {\em Gen.Rel.Grav.}, 34:1445, 2002.

\bibitem{ellis02}
G.F.R. Ellis, P.~McEwan, W.~Stoeger, and P.~Dunsby.
\newblock Causality of inflationary universes with positive spatial curvature.
\newblock {\em Gen.Rel.Grav.}, 34:1461, 2002.

\bibitem{phisky}
A.N. Lasenby and C.~Doran.
\newblock Conformal models of de {S}itter space and initial conditions for
  inflation.
\newblock To appear in proceedings `Phi in the Sky', Porto, astro-ph/0411579,
  2004.

\bibitem{anl-confcos}
A.N. Lasenby.
\newblock Conformal geometry and the universe.
\newblock \textit{Phil. Trans. R. Soc. Lond. A}, to appear, 2003.

\bibitem{rat95}
B.~Ratra and P.J.E. Peebles.
\newblock Inflation in an open universe.
\newblock {\em Phys. Rev. D}, 52(4):1837--1894, 1995.

\bibitem{che68}
N.A. Chernikov and E.A. Tagirov.
\newblock Quantum theory of scalar fields in de {S}itter space-time.
\newblock {\em Ann. Inst. Henri Poincar{\'{e}}}, 9A:109, 1968.

\bibitem{bir-quant}
N.D. Birrell and P.~C.~W. Davies.
\newblock {\em Quantum Fields in Curved Space}.
\newblock Cambridge University Press, 1982.

\bibitem{all85}
B.~Allen.
\newblock Vacuum states in de {S}itter space.
\newblock {\em Phys. Rev. D}, 32:3136, 1985.

\bibitem{dan02}
U.H. Danielsson.
\newblock Inflation, holography, and the choice of vacuum in de {S}itter space.
\newblock {\em J. High Energy Phys.}, 07:40, 2002.

\bibitem{gre-rqm}
W.~Greiner.
\newblock {\em Relativistic Quantum Mechanics}.
\newblock Springer--Verlag, Berlin, 1990.

\bibitem{brand01}
R.H. Brandenberger.
\newblock Principles, progress and problems in inflationary cosmology.
\newblock {\em AAPPS Bull.}, 11:20, 2001.

\bibitem{cor98}
N.~Cornish and E.P.S. Shellard.
\newblock Chaos in quantum cosmology.
\newblock {\em Phys. Rev. Lett.}, 81:3571, 1998.

\bibitem{LLinf}
A.R. Liddle and D.H. Lyth.
\newblock {\em Cosmological Inflation and Large-Scale Structure}.
\newblock Cambridge University Press, 2000.

\bibitem{mar00}
J.~Martin and D.J. Schwarz.
\newblock The precision of slow-roll predictions for the {CMBR} anisotropies.
\newblock {\em Phys.Rev.}, D62:103520, 2000.

\bibitem{efs03}
G.~Efstathiou.
\newblock Is the low {CMB} quadrupole a signature of spatial curvature?
\newblock {\em MNRAS}, 343:L95, 2003.

\bibitem{fran03}
G.~Franco, P.~Fosalba, and J.A. Tauber.
\newblock Systematic effects in the measurement of polarization by the {PLANCK}
  telescope.
\newblock {\em Astron. Astrophys.}, 405:349, 2003.

\bibitem{muk92}
V.~F. Mukhanov, H.~A. Feldman, and R.~H. Brandenberger.
\newblock Theory of cosmological perturbations.
\newblock {\em Phys. Rep.}, 215:203--333, 1992.

\bibitem{zhang03}
D.~Zhang and C.~Sun.
\newblock The exact evolution equation of the curvature perturbation for closed
  universe.
\newblock astro-ph/0310127, 2003.

\bibitem{par69}
L.~Parker.
\newblock Quantized fields and particle creation in expanding universes. {I}.
\newblock {\em Phys. Rev}, 183:1057--1068, 1969.

\bibitem{ful-qft}
S.A. Fulling.
\newblock {\em Aspects of Quantum Field Theory in Curved Space-Time}.
\newblock Cambridge University Press, 1989.

\bibitem{CAMB}
A.M. Lewis, A.D. Challinor, and A.N. Lasenby.
\newblock Efficient computation of {CMB} anisotropies in closed {FRW} models.
\newblock {\em Ap. J.}, 538:473, 2000.

\bibitem{free01}
W.L.~Freedman et~al.
\newblock Final results from the {H}ubble space telescope key project to
  measure the {H}ubble constant.
\newblock {\em Ap. J.}, 553:47, 2001.

\bibitem{jones01}
M.E.~Jones et~al.
\newblock {$H_0$} from an orientation-unbiased sample of {SZ} and {X}-ray
  clusters.
\newblock astro-ph/0103046, 2001.

\bibitem{perc01}
W.J.~Percival et~al.
\newblock The {2dF} galaxy redshift survey: The power spectrum and the matter
  content of the universe.
\newblock {\em MNRAS}, 327:1297, 2001.

\bibitem{eis98}
D.J. Eisenstein and W.~Hu.
\newblock Baryonic features in the matter transfer function.
\newblock {\em Ap. J.}, 496:605, 1998.

\bibitem{vsa04}
{C. Dickinson et al.}
\newblock High sensitivity measurements of the {CMB} power spectrum with the
  extended {V}ery {S}mall {A}rray.
\newblock astro-ph/0402498, 2004.

\bibitem{page84}
D.N. Page.
\newblock A fractal set of perpetually bouncing universes.
\newblock {\em Class. Quant. Grav.}, 1:417, 1984.

\bibitem{efs03a}
G.~Efstathiou.
\newblock The statistical signficance of the low {CMB} multipoles.
\newblock {\em {MNRAS}}, 346:L26, 2003.

\bibitem{efs04}
G.~Efstathiou.
\newblock A maximum likelihood analysis of the low {CMB} multipoles from
  {WMAP}.
\newblock {\em MNRAS}, 348:885, 2004.

\bibitem{cont03}
C.R. Contaldi, M.~Peloso, L.~Kofman, and A.~Linde.
\newblock Suppressing the lower multipoles in the {CMB} anisotropies.
\newblock {\em JCAP}, 07(002):1, 2003.

\end{thebibliography}

\end{document}